\documentclass[nofootinbib,reprint,floatfix, longbibliography, superscriptaddress]{revtex4-1}

\usepackage{float}
\usepackage[utf8]{inputenc}
\usepackage{amsmath}
\usepackage{amssymb}
\usepackage{graphicx}
\usepackage{verbatim}
\usepackage{color}
\usepackage{subfigure}
\usepackage{notes2bib}
\usepackage{comment}
\usepackage{hyperref}
\usepackage{xcolor}
\hypersetup{colorlinks, linkcolor={red!50!black},citecolor={blue!50!black},urlcolor={blue!80!black}}
\usepackage{physics}

\usepackage{epstopdf}

\newcommand*\diff{\mathop{}\!\mathrm{d}}

\begin{document}

\title{Faithful conversion of propagating quantum information to mechanical motion}

\author{A. P. Reed}
\affiliation{JILA, Boulder, Colorado 80309-0440, USA}
\affiliation{Department of Physics, University of Colorado, Boulder, CO 80309-0390, USA}

\author{K. H. Mayer}
\affiliation{Department of Physics, University of Colorado, Boulder, CO 80309-0390, USA}
\affiliation{National Institute of Standards and Technology (NIST), Boulder, CO 80305, USA}

\author{J. D. Teufel}
\affiliation{National Institute of Standards and Technology (NIST), Boulder, CO 80305, USA}

\author{L. D. Burkhart}
\affiliation{Departments of Applied Physics and Physics, Yale University, New Haven, CT 06520, USA}

\author{W. Pfaff}
\affiliation{Departments of Applied Physics and Physics, Yale University, New Haven, CT 06520, USA}

\author{M. Reagor}
\affiliation{Departments of Applied Physics and Physics, Yale University, New Haven, CT 06520, USA}
\affiliation{Rigetti Computing, 775 Heinz Avenue, Berkeley, CA 94710, USA}

\author{L. Sletten}
\affiliation{JILA, Boulder, Colorado 80309-0440, USA}
\affiliation{Department of Physics, University of Colorado, Boulder, CO 80309-0390, USA}

\author{X. Ma}
\affiliation{JILA, Boulder, Colorado 80309-0440, USA}
\affiliation{Department of Physics, University of Colorado, Boulder, CO 80309-0390, USA}

\author{R. J. Schoelkopf}
\affiliation{Departments of Applied Physics and Physics, Yale University, New Haven, CT 06520, USA}

\author{E. Knill}
\affiliation{National Institute of Standards and Technology (NIST), Boulder, CO 80305, USA}
\affiliation{Center for Theory of Quantum Matter, University of Colorado, Boulder, CO 80309, USA}

\author{K. W. Lehnert}
\affiliation{JILA, Boulder, Colorado 80309-0440, USA}
\affiliation{Department of Physics, University of Colorado, Boulder, CO 80309-0390, USA}
\affiliation{National Institute of Standards and Technology (NIST), Boulder, CO 80305, USA}

\date{\today}

\begin{abstract}
We convert propagating qubits encoded as superpositions of zero and one photons to the motion of a micrometer-sized mechanical resonator. Using quantum state tomography, we determine the density matrix of both the propagating photons and the mechanical resonator. By comparing a sufficient set of states before and after conversion, we determine the average process fidelity to be $F_{\textrm{avg}} = 0.83\substack{+0.03\\-0.06}$ which exceeds the classical bound for the conversion of an arbitrary qubit state. This conversion ability is necessary for using mechanical resonators in emerging quantum communication and modular quantum computation architectures.
\end{abstract}

\maketitle

%%%%%%%%%%%%%%%%%%%%%%%%%%%%%%%%%%%%%%%%%%%%%%%%%%%
% Main Text
%%%%%%%%%%%%%%%%%%%%%%%%%%%%%%%%%%%%%%%%%%%%%%%%%%%
The motion of micrometer-sized mechanical resonators can now be controlled and measured at the fundamental limits imposed by quantum mechanics.  Such resonators have been prepared in their motional ground state\cite{OConnell2010, Teufel2011, Chan2011} or in squeezed states\cite{Wollman2015,Lecocq2015b, Pirkkalainen2015}, measured with quantum limited precision\cite{Teufel2009,Anetsberger2010}, and even entangled with microwave fields\cite{Palomaki2013b}. Beyond these fundamental advances, mechanical resonators are emerging as potential high-fidelity interfaces for quantum information between the microwave and optical domains\cite{Bochmann2013, Bagci2014, Andrews2014, Fang2016} and as on-demand memory elements for superconducting quantum circuits\cite{Palomaki2013a}. Such capabilities have direct applications in recently proposed quantum communication and modular quantum computation architectures\cite{Monroe2014,Narla2016}.

Specifically, quantum communication networks that use superconducting qubits and modular quantum computing architectures require the ability to store, amplify, or frequency-shift propagating microwave fields. A single electromechanical device provides all of these functions by rapidly varying the parametric coupling between mechanical motion and microwave fields\cite{Palomaki2013a,Andrews2015,Lecocq2016,Ockeloen-Korppi2016}. For example, the ability to suddenly turn off the interaction between a microwave field and mechanical motion allows the state of a field propagating through a transmission line to be converted to, and trapped in, the motional state of the resonator\cite{Palomaki2013a}.

To use the conversion process as part of a general quantum information processor, one must work with states that have non-Gaussian statistics, such as qubits encoded as superpositions of zero and one photons. In contrast, any process using only Gaussian states can be simulated efficiently on a classical computer\cite{Bartlett2002}. But most electromechanical devices operate in a regime in which the equations that describe the coupling are linear, ensuring that a Gaussian state of the microwave field or mechanical resonator will never evolve into a non-Gaussian state. Accessing non-Gaussian mechanical states requires either a source of non-Gaussian microwave fields\cite{OConnell2010} or a nonlinear detector such as a single photon counter\cite{Lecocq2015a, Riedinger2016}.

\begin{figure}[!b]
  \centering
    \includegraphics[scale=0.96]{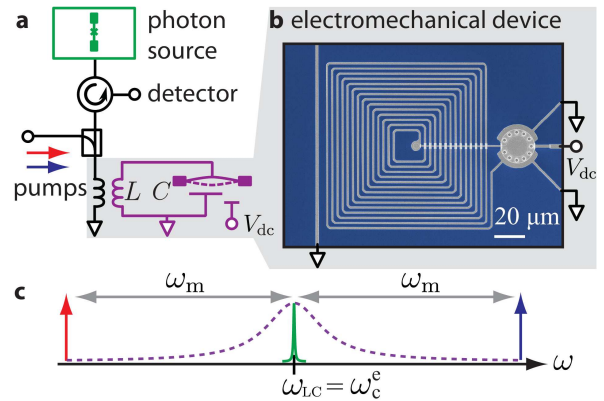}
      \caption{Diagram of the experiment. \textbf{a,} A simplified schematic shows the electromechanical device connected to a photon source consisting of a transmon qubit in a microwave cavity. Pumps (arrows) used to create the capture (red) and amplification (blue) interactions are injected into the transmission line. \textbf{b,} False-color micrograph of the electromechanical device where aluminium films (gray) are deposited on a sapphire substrate (blue). The spiral forms the inductor and the disc is the mechanically-compliant capacitor. \textbf{c,} Pumps are detuned below and above the LC circuit's resonant frequency $\omega_{\textrm{LC}}$ by the mechanical resonant frequency $\omega_{\textrm{m}}$. Using a voltage bias $V_{\textrm{dc}}$, the LC circuit's resonant response (dashed magenta) is tuned to match the much narrower resonance of the microwave cavity (green solid) at the frequency $\omega_{\textrm{c}}^{\textrm{e}}$.}
\label{fig:pitchCatchSetup}
\end{figure}

In this work, we convert non-Gaussian states from propagating microwave fields to the motion of a micrometer-sized mechanical resonator. We use an electromechanical device to capture, store, and amplify single photons generated by a superconducting qubit and then determine the density matrix of the mechanical resonator using quantum state tomography. We find that the quantum state can be stored for a characteristic time that \mbox{exceeds 100~$\mu$s,} an improvement of over four orders of magnitude compared to previous work that demonstrated the storage of a non-Gaussian state in an electromechanical device\cite{OConnell2010}. To characterize how the \mbox{capture~\textit{process}} affects arbitrary propagating qubit states, we capture superpositions of zero and one photons. The degree to which this process preserves quantum information is quantified by the average fidelity\cite{Nielsen2002}, which we find to be $F_{\textrm{avg}} = 0.83\substack{+0.03\\-0.06}$ where the limits are the 90\% confidence interval. This level of performance exceeds the fidelity achievable using only classical resources, indicating that our electromechanical device is suitable for the transduction of quantum information.

The electromechanical device consists of an inductor-capacitor (LC) circuit that is tunable and coupled to a mechanical resonator (Fig.~\ref{fig:pitchCatchSetup}a). The tunability and coupling arise from the compliant upper plate of the parallel-plate capacitor, which is a 100~nm thick suspended and tensioned aluminium membrane that is free to vibrate. The fundamental drumhead-like vibrational mode of this membrane forms the mechanical resonance at $\omega_{\textrm{m}}/2\pi \approx 9.3 \textrm{ MHz}$. If displaced by the resonator's zero-point motion of 6.4~fm, the circuit's resonant frequency shifts by $g_0/2\pi \approx 280$~Hz. The circuit also couples inductively to propagating microwave fields in a nearby transmission line at a rate of $\kappa_{\textrm{LC}}/2\pi~\approx~3~\textrm{ MHz}$. We tune the LC circuit into precise resonance with a narrow-band and fixed-frequency photon source by using a third electrode, biased at $V_{\textrm{dc}}$ relative to the membrane, to control the static separation between the membrane and the microwave electrode\cite{Andrews2015}.

We connect the electromechanical device to an on-demand source of single photons using the network depicted in Fig.~\ref{fig:pitchCatchSetup}a. To efficiently generate single photons compatible with the narrow bandwidth requirements\cite{Palomaki2013a} of the electromechanical device, we use a circuit quantum electrodynamics (cQED) system\cite{Paik2011}. It consists of a transmon qubit with a transition frequency $\omega_{\textrm{q}}/2\pi = 5.652$~GHz in a microwave cavity, whose resonance frequency is $\omega_{\textrm{c}}^g/2\pi = 7.290 \textrm{ GHz}$ when the qubit is in the ground state, $|g\rangle$, and $\omega_{\textrm{c}}^e= \omega_{\textrm{c}}^g - 2\chi$ = 7.283 GHz when it is in the excited state, $|e\rangle$, where $\chi$ is the dispersive shift. We use a control pulse\cite{Kindel2016} to drive the transition $|g\rangle|0\rangle \rightarrow|e\rangle|1\rangle$ where $|0\rangle$ and $|1\rangle$ correspond to zero and one cavity photons, respectively. The cavity state then evolves into a field propagating through the transmission line with the center frequency~$\omega_{\textrm{c}}^e$ and narrow bandwidth~$\kappa_{\textrm{c}}/2\pi =~60$~kHz.

\begin{figure}[!ht]
  \centering
    \includegraphics[scale=0.88]{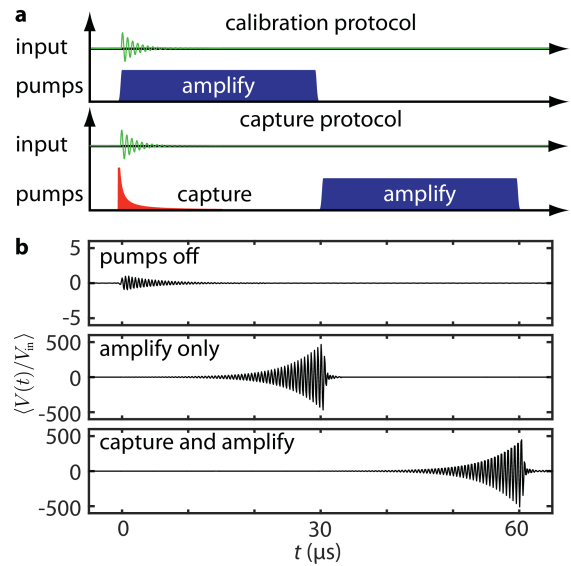}
      \caption{Calibration and capture protocols. \textbf{a,} Timing diagrams depicting the input coherent signal (green) of amplitude $V_{\textrm{in}}$ and the pulse shapes of the pumps, $\Gamma_{\textrm{r}}(t)$ and $\Gamma_{\textrm{b}}(t)$, used to create the capture or amplification interaction. For the calibration protocol (top), the amplification pump is coincident with the signal pulse. The capture protocol (bottom) has a timing diagram similar to the calibration protocol, but the input signal is coincident with a capture pulse that is temporally shaped for optimal capture of the signal. At $t = 30$~$\mu$s, the mechanical state is amplified and converted back into a microwave field. \textbf{b,} The plots show the voltage signals, $V(t)$, measured at the detector and averaged over 500 repetitions of each protocol when the pumps were either off or on. During amplification, $\Gamma_{\textrm{b}}/2\pi = 60$ kHz which results in a gain of 53~dB.}
\label{fig:classicalSignals}
\end{figure}

\begin{figure*}[!ht]
  \centering
    \includegraphics[scale=0.86]{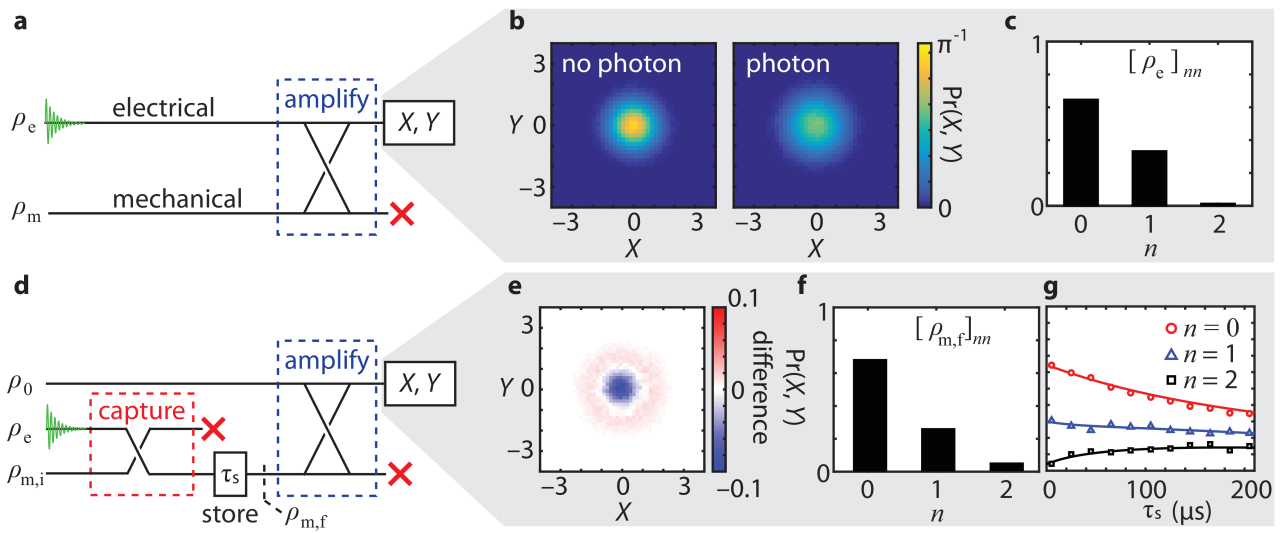}
      \caption{Capture, storage, and amplification of single propagating photons. \textbf{a,} The diagram shows the calibration protocol where the green decaying sinusoid represents an input microwave field in the state $\rho_{\textrm{e}}$. Prior to injecting an input state, the mechanical mode described by $\rho_{\textrm{m}}$ is cooled close to its quantum ground state. A red cross indicates a mode was not measured. \textbf{b,} Histograms of 512,000 measured quadrature amplitudes $X$ and $Y$ for the input state are plotted such that the histograms form a discretized and normalized joint probability distribution $\textrm{Pr}(X, Y)$. The labels `no photon' and `photon' indicate whether single photons were generated or not. \textbf{c,} The diagonal elements of $\rho_{\textrm{e}}$ are obtained using a maximum likelihood state tomography (Supplemental Information). \textbf{d,} The diagram shows the capture protocol where $\rho_{\textrm{m,i}}$ is the initial mechanical state and $\tau_{\textrm{s}}$ is an adjustable storage time. After storage, the mechanical mode is in the final state $\rho_{\textrm{m,f}}$. During amplification, the electrical mode is in a vacuum state $\rho_{0}$. \textbf{e,} The figure shows the difference of the `photon' and `no photon' histograms acquired using the capture protocol, highlighting the phase symmetric character of a single \textit{phonon} state. \textbf{f,} Diagonal elements of $\rho_{\textrm{m}}$ obtained at $\tau_{\textrm{s}} < 3$ $\mu$s. \textbf{g,} The diagonal elements of $\rho_{\textrm{m}}$ decay toward their thermal equilibrium values as a function of $\tau_{\textrm{s}}$. A model (solid lines) of $\rho_{\textrm{m}}$ yields a characteristic storage time of $\tau_{\textrm{m}} = 137\pm6$~$\mu$s (Supplemental Information).} 
\label{fig:photonData}
\end{figure*}

% paragarphs about the two protocols
The propagating microwave field is parametrically coupled to the membrane's motion by applying pumps to the LC circuit. To capture the state of the propagating field\cite{Palomaki2013a}, we use a pump that is detuned below (red-detuned) the LC resonance with detuning $\Delta_\textrm{r}~= -\omega_{\textrm{m}}.$ This pump implements a beamsplitter interaction that swaps the states of the input microwave field and the mechanical resonator. For a given temporal envelope of the input field, the coupling rate between the resonator and the field, $\Gamma_\textrm{r}(t)~=~4 g_0^2 n_{\textrm{r}}(t) / \kappa_{\textrm{LC}}$, must be modulated for optimal capture efficiency\cite{Andrews2015}, where $n_{\textrm{r}}(t)$ is the number of photons induced in the LC circuit by the pump. If instead we apply a blue-detuned pump at $\Delta_\textrm{b}~= +\omega_{\textrm{m}}$, a two-mode squeezer interaction is created that amplifies both the motion of the resonator and the incident microwave field\cite{Palomaki2013b}. During amplification, the LC circuit emits a propagating field with a temporal envelope that rises exponentially at a rate of $\Gamma_\textrm{b}/2$ where $\Gamma_\textrm{b}(t)~=  4 g_0^2 n_{\textrm{b}}(t) / \kappa_{\textrm{LC}}$ is set by the strength,~$n_{\textrm{b}}(t)$, of the blue-detuned pump. Crucially, the state of the emitted field depends on both the states of the resonator and the input field before amplification\cite{Caves2012}.

We exploit the parametric interactions in two protocols that are used to characterize the capture process. Because the process maps states at the input of the electromechanical device to the resonator, we must determine the input state and compare it to the captured state. To this end, we have developed `calibration' and `capture' protocols that enable us to determine the input and captured states, respectively (Fig.~\ref{fig:classicalSignals}a). We initially test the two protocols with coherent signals whose frequency and bandwidth are chosen to match those created by the cQED system~(Fig.~\ref{fig:classicalSignals}b).

In the calibration protocol, the input field is amplified directly and then measured. We implement it by applying the blue-detuned pump coincident with the input field. In this case, the electromechanical device functions as a linear phase-preserving amplifier whose input and output are the incident and reflected microwave fields, respectively. These pulsed fields have different envelopes; nevertheless, with an appropriate filter (Supplementary Information) they are related by an energy gain of $\cosh^2 \left(r/2 \right)$ where $r=\Gamma_{\textrm{b}}\tau_{\textrm{b}}$ and $\tau_{\textrm{b}}$ is the pump's duration. If we regard the input of the amplifier as the incident microwave field, the fluctuations of the resonator's motion are the source of the amplifier's added noise, reaching the quantum limit\cite{Caves2012} if the resonator is in its ground state\cite{Teufel2011}.

After obtaining the input state, we use the capture protocol to determine the resonator state. We first apply the red-detuned pump coincident with the input field. Once it is captured, we then apply the blue-detuned pump to amplify the resonator's state. In contrast to the calibration protocol, we now regard the amplifier's input to be the state of the resonator. The output is still the reflected field, but the added noise is due to the vacuum fluctuations of the incident field. When interpreted this way, we realize a linear phase-conjugating amplifier with an energy gain of~$\sinh^2 \left(r/2 \right)$.

Operating the electromechanical device as a low-noise amplifier enables us to perform state tomography on both the input microwave field and on the motion of the resonator. For each repetition of the two protocols depicted in Fig.~\ref{fig:photonData}, we record a voltage signal,~$V(t)$, at the detector during the amplification. For each voltage record, we extract a pair of quadrature amplitudes,~$X$ and $Y$, for the state of either the resonator or input field (Supplementary Information). By making repeated measurements of $V(t)$, we obtain a set of quadrature amplitudes and use this information to extract a density matrix $\rho$ via a method of maximum likelihood state tomography\cite{Eichler2011} (Supplemental Information). We refer to the states of the input microwave field and of the mechanical resonator as $\rho_{\textrm{e}}$ and $\rho_{\textrm{m}}$, respectively.

\begin{figure*}[!ht]
  \centering
    \includegraphics[scale=0.78]{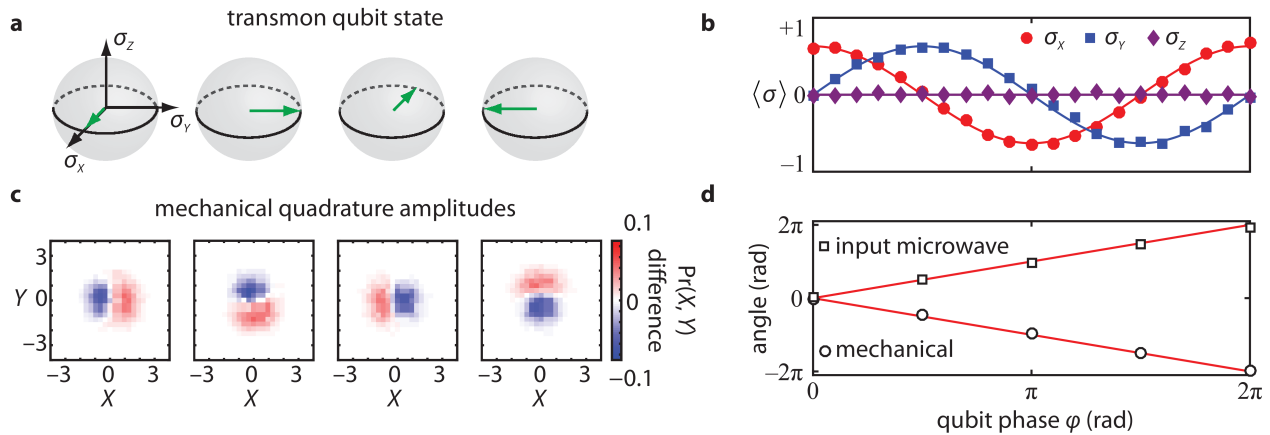}
      \caption{Conversion of propagating qubits. \textbf{a,} The transmon qubit was prepared in the superposition state ${1 \over \sqrt{2}} \left( |g\rangle + e^{i \varphi} |e\rangle \right)$ with a phase $\varphi$ chosen from the set $\{0, \pi/2, \pi, 3\pi/2\}$, as denoted by the green arrow plotted on a Bloch sphere. \textbf{b,} The plot shows measured Pauli component amplitudes, $\langle \sigma_{k} \rangle$, of the qubit state as a function of its phase where $k=\{x,y,z\}$. Single-shot readout of the qubit state was achieved by using the electromechanical device as a nearly quantum-limited amplifier. The solid lines are fits with the readout contrast of $60\%$ as the only free parameter. \textbf{c,} Subtracted histograms (similar to Fig.~\ref{fig:photonData}e) show the mechanical quadrature amplitudes, $X$ and $Y$, as the transmon qubit's phase was varied. \textbf{d,} The plot shows the argument of the off-diagonal density matrix element, $\rho_{01}$, for both the input microwave, $\rho_{\textrm{e}}$, and captured mechanical, $\rho_{\textrm{m}}$, states as a function of $\varphi$. The mechanical state changes linearly in $\varphi$, indicating that the conversion process is coherent. The apparent opposite dependence of $\varphi$ of the input and captured states is a result of the phase conjugate amplification of the mechanical state compared to the direct amplification of the input microwave state (solid lines indicate the expected behavior).}
\label{fig:qubitData}
\end{figure*}

To test the conversion of non-Gaussian states, we inject single photons into the electromechanical device. We also operate the calibration and capture protocols (Figs.~\ref{fig:photonData}a,d) without generating single photons to determine the gain of the detector which we use to scale $X$ and $Y$ in units of $(\textrm{quanta})^{1/2}$ (Supplemental Information). Prior to the execution of each protocol, we cool the resonator close to its quantum ground state\cite{Teufel2011} with an occupancy of approximately 0.1~quanta (Supplemental Information). For both protocols, the tomography yields density matrix estimates containing significant elements only on the diagonals (Fig.~\ref{fig:photonData}c). In particular, we find
that the probability of detecting a single photon is $\left[ \rho_{\textrm{e}}\right]_{11} = 0.33\substack{+0.02 \\ -0.01}$ (Supplemental Information). After capture, the probability of a single \textit{phonon} occupying the mechanical mode is $\left[\rho_{\textrm{m}}\right]_{11} = 0.26\substack{+0.01 \\ -0.02}$. To distinguish the captured state from a thermal or coherent state, we calculate the degree of second-order coherence $g^{(2)}_{\textrm{m}}~= 0.89\substack{+0.05 \\ -0.17}$ (Supplemental Information). For comparison, a thermal or coherent state of motion yields $g^{(2)}_{\textrm{m}} \geq 1$. After capturing the mixed single photon state, we vary the storage time~$\tau_{\textrm{s}}$ and test the ability to mechanically store a non-Gaussian state (Fig.~\ref{fig:photonData}f). 
We use a master equation formalism to model the evolution of $\rho_{\textrm{m}}$ with the characteristic storage time $\tau_{\textrm{m}}$ as the only free parameter (Supplemental Information). We extract $\tau_{\textrm{m}} = 137\pm6$~$\mu$s, which is about ten times longer than the time used to capture the input photon state.

Having demonstrated the ability to capture single photons, we then characterize how the capture process affects arbitrary qubit states encoded as superpositions of zero and one photons. This process is described by a map~$\mathcal{E}$ between incident and captured states whose quality is characterized by the average fidelity\cite{Nielsen2002}
\begin{equation}
F_{\mathrm{avg}}=~\int d\Psi\bra{\Psi}\mathcal{E}(\ket{\Psi}\bra{\Psi})\ket{\Psi},
\end{equation}
which measures how indistinguishable the output of the process is from the input, averaged over all pure input states~$\ket{\Psi}$. To determine $F_{\mathrm{avg}}$, it is sufficient to capture a set of states that includes a single photon state and superpositions of zero and one photons. We can create superposition states by first preparing the transmon qubit in the superposition ${1 \over \sqrt{2}} \left( |g\rangle + e^{i \varphi} |e\rangle \right)$, with varying phase $\varphi$, as shown in Fig.~\ref{fig:qubitData}a. By driving the transition $|g\rangle|0\rangle \rightarrow|e\rangle|1\rangle$, we transfer the superposition state from the transmon to the cavity and then let the cavity state evolve into the propagating field. Operating the capture protocol on this set of states shows that the phase of the qubit state is converted to the motion of the mechanical resonator (Fig.~\ref{fig:qubitData}c). More quantitatively, we follow the procedure illustrated in Fig.~\ref{fig:photonData}, determining both $\rho_{\textrm{e}}$ and $\rho_{\textrm{m}}$ for this set of states (Fig.~\ref{fig:qubitData}d). The input and captured density matrices provide enough information to calculate $F_{\textrm{avg}} = 0.83\substack{+0.03\\-0.06}$ for arbitrary qubit states. Crucially, the average fidelity exceeds~2/3, the highest possible fidelity for transferring qubits using only classical resources~(Supplemental~Information).

Converting microwave qubit states to mechanical motion opens up new possibilities to process quantum information using micrometer-sized mechanical resonators. To communicate quantum information between remote modules in a network, such resonators may be the key element in the transduction of microwave quantum signals to telecommunications light\cite{Bochmann2013, Bagci2014, Andrews2014, Fang2016}. The communication can be made robust against transmission loss by transducing multiphoton states\cite{Pfaff2016}. For quantum computation protocols that require the feed-forward of information, such as teleportation\cite{Bennett1993} and error correction schemes\cite{Ofek2016}, mechanical resonators can act as on-demand memories for quantum states. As microfabrication advances continue to reduce mechanical dissipation, it will become possible to store a quantum state in the motion of a macroscopic object on the minute timescale\cite{Norte2016,Reinhardt2016}.

\subsection*{Acknowledgments}
We acknowledge advice from C. Axline, M. Castellanos-Beltran, L. Frunzio, S. Glancy, W. F. Kindel, and F. Lecocq as well as technical assistance from R. Delaney and H. Greene. We thank P. Blanchard for taking the micrograph shown in Fig.~1. We acknowledge funding from National Science Foundation (NSF) under Grant Number 1125844, AFOSR MURI under grant number FA9550-15-1-0015, and the Gordon and Betty Moore Foundation. A. P. R. acknowledges support from the NSF Graduate Research Fellowship under grant number DGE 1144083. L. D. B. acknowledges the support of the ARO QuaCGR Fellowship. This is a contribution of NIST, an agency of the US government, not subject to copyright.


%%%%%%%%%%%%%%%%%%%%%%%%%%%%%%%%%%%%%%%%%%%%%%%%%%%
% Supplemental Information
%%%%%%%%%%%%%%%%%%%%%%%%%%%%%%%%%%%%%%%%%%%%%%%%%%%

%\clearpage

\onecolumngrid
\newpage
\appendix
\renewcommand\thefigure{\thesection\arabic{figure}}    
\setcounter{figure}{0}    
\renewcommand\thetable{S\arabic{table}}    
\setcounter{table}{0}    
\setcounter{equation}{0}
\setcounter{section}{19}
\section*{Supplementary Information}

\subsection{Measurement network}

\begin{figure}[!ht]
  \centering
    \includegraphics[scale=0.8]{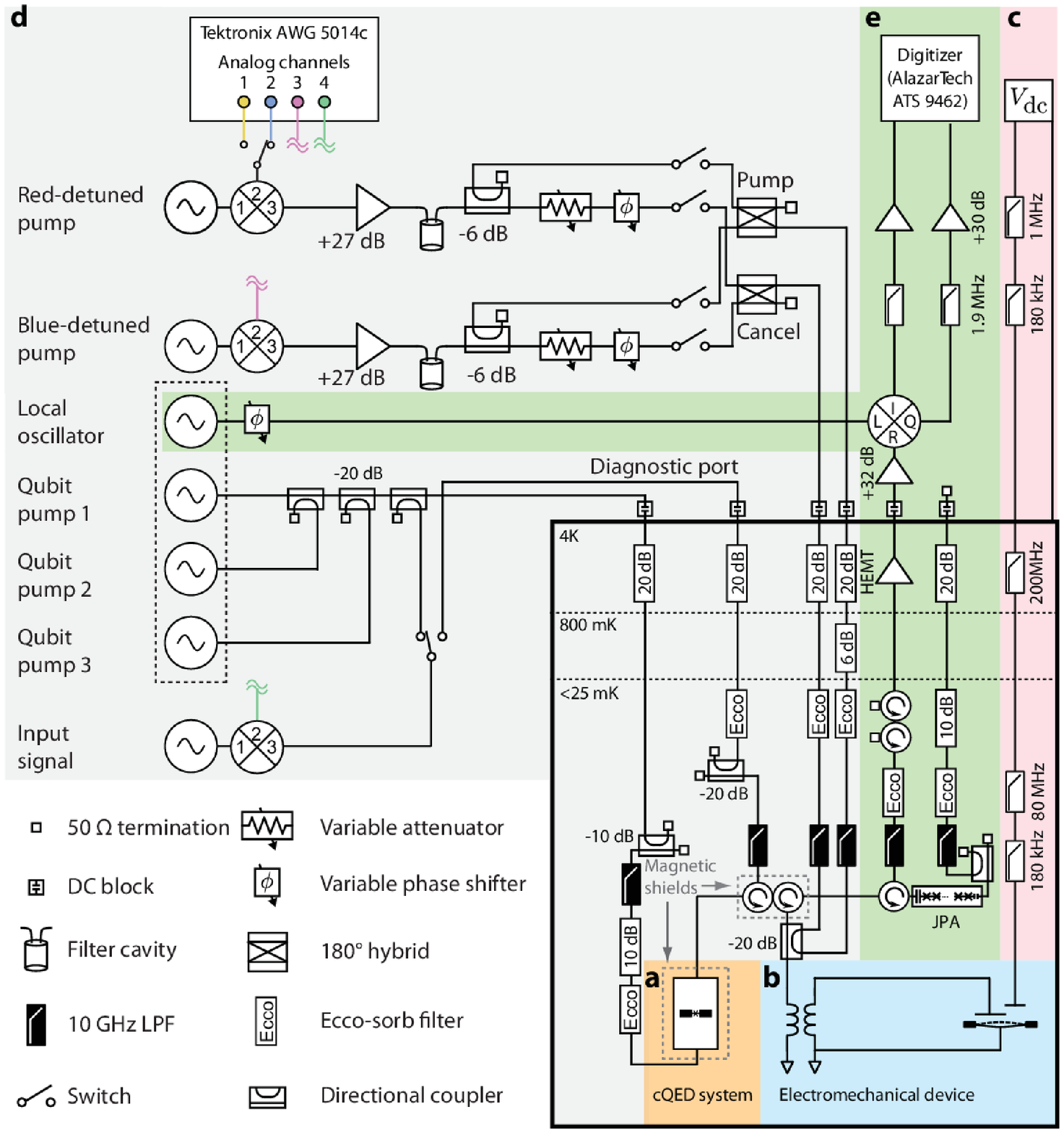}
      \caption{Measurement network. \textbf{a,} Circuit quantum electrodynamics (cQED) system. \textbf{b,} Electromechanical device. \textbf{c,} Actuation line. \textbf{d} Microwave pumps and signal synthesis. \textbf{e,} Detector.}
      
\label{fig:QNetworkSchem}
\end{figure}

Supplementary~Fig.~\ref{fig:QNetworkSchem} shows a diagram of the measurement network. The network consists of five main parts: A transmon qubit is embedded in a microwave cavity (cQED system) with two ports (Supplementary~Fig.~\ref{fig:QNetworkSchem}a). The cQED system is connected to an electromechanical device, which is mounted to the base stage of a dilution refrigerator (Supplementary~Fig.~\ref{fig:QNetworkSchem}b). The center frequency of the electromechanical device is controlled by a voltage bias provided by an actuation line (Supplementary~Fig.~\ref{fig:QNetworkSchem}c). Microwave pumps and test signals are routed to both the cQED system and the electromechanical device (Supplementary~Fig.~\ref{fig:QNetworkSchem}d). Microwave signals are measured using a detector (Supplementary~Fig.~\ref{fig:QNetworkSchem}e).

\subsubsection{Electromechanical device}
The electromechanical device is mounted to the mixing chamber of a dilution refrigerator and cooled to $<25$ mK. The construction and operation of the electromechanical device is described in Ref. \cite{Andrews2015}.

\subsubsection{Transmon qubit and microwave cavity}
The transmon qubit consists of an Al/AlO$_x$/Al Josephson junction shunted by a superconducting aluminum coplanar capacitor. This circuit is lithographically fabricated on a single-crystal sapphire substrate and the coplanar capacitor acts as a dipole antenna.

%At room temperature, the junction has a normal state resistance of 8.8 k$\Omega$.

We embed the transmon qubit in a three-dimensional microwave cavity, as described in Ref. \cite{Paik2011}. The cavity is milled from a two pieces of extruded T6061 aluminium and the inner surfaces of the cavity walls are mechanically polished. Two holes that serve as microwave coupling ports are drilled into the cavity. One of the ports is weakly coupled to a pump line, which we use to excite either the qubit or cavity. The other cavity port is strongly coupled to a transmission line which is routed to components connected to the electromechanical device.

To reduce stray magnetic fields that can affect the qubit's coherence times, we use non-magnetic materials such as copper and brass to construct components that are in close proximity to the qubit and cavity. For additional magnetic shielding, we enclose the cavity in a Cryoperm magnetic shield (Amuneal Manufacturing). At room temperature and inside the magnetic shield, we measure an ambient magnetic field of $\sim20$ mG near the cavity. During operation of the experiment, the magnetic shield is wrapped in absorptive microwave material (Eccosorb) and a thin layer (\mbox{$<100$ $\mu$m}) of aluminium.

Signals emitted from the cavity are routed using copper cables to the input of two low insertion loss circulators and a directional coupler. These custom cryogentic and magnetically shielded circulators (Raditek, Inc.) are used to route signals emitted from the cavity to the electromechanical device, while also providing isolation from the high power ($<10$~nW) red- and blue-detuned pumps (Supplementary~Section~\ref{sec:electromechanicsEOM}). At 300 K and near 7.283 GHz, the isolation of the two circulators connected in series was measured to be $-43$ dB. A similar level of isolation was measured at 4 K. Each circulator is specified by the manufacturer to have an insertion loss of 0.2 dB (at 100 mK). Following the circulators, the output of the directional coupler is connected to the electromechanical device using a superconducting niobium-titanium (NbTi) cable.

\subsubsection{Actuation line}
The voltage on the actuation line controls the center frequency of the electromechanical device. A stable and low noise voltage source (Yokogawa 7651) provides a constant $V_{\textrm{dc}}$ during the operation of the experiment. Filtering on the actuation line is nearly identical to the configuration described in Ref. \cite{Andrews2015}.

\subsubsection{Arbitrary microwave pump and signal generation}

The red- and blue-detuned pumps at $\omega_{\mathrm{LC}}-\omega_{\mathrm{m}}$ and $\omega_{\mathrm{LC}}+\omega_{\mathrm{m}}$, respectively, are generated using two separate microwave synthesizers (Agilent PSG). However, these synthesizers by themselves cannot produce microwave pulses with programmable temporal envelopes as required by the protocols depicted in Fig.~2 of the main text. To generate such pulses, we shape the temporal envelopes of microwave tones emitted by the synthesizers. To this end, we use a double-balanced mixer (Marki MM1-0625HS) driven by waveforms with a programmable amplitude provided by an arbitrary waveform generator (Tektronix AWG 5014c). Waveforms generated by the AWG are shown in Supplementary Fig.~\ref{fig:TimingDiagram}. The shaped microwave pulses have a dynamic range of approximately 50 dB, set by the LO-RF isolation of the double-balanced mixer. Additional isolation (80 dB) is achieved by pulsing off the microwave synthesizers when the pumps are not needed. The shaped pulses are then amplified (Mini-Circuits ZVA-183V) and filtered\cite{Rogers1993}. Additionally, the pumps have Gaussian-smoothed edges given by a characteristic time of $\sigma_t>200$~ns. Such smoothed edges reduce spectral content at $\omega_{\mathrm{LC}}\pm\omega_{\mathrm{m}}$ that could drive the mechanical resonator.

The red- and blue-detuned pumps carry enough power that could adversely affect the qubit's state. To reduce the pump power incident on the cQED system, we use variable attenuators and phase shifters to create cancellation signals that reduce the pump power incident on the cQED system by 30 to 40 dB. We monitor and adjust the relative cancellation of the pumps at the detector.

To test the calibration and capture protocols, we inject a large amplitude coherent signal into the electromechanical device (shown in Fig.~2 of the main text). This test signal is generated using a microwave synthesizer, and then shaped using a double-balanced mixer. It's temporal envelope is shown in Supplementary~Fig.~\ref{fig:TimingDiagram}. While executing the protocols for the experiments depicted in Figs.~3 and 4 of the main text, the test signal was not injected into the network.

Qubit pumps and the experiment's local oscillator (LO) are generated using a set of phase coherent microwave synthesizers (Holzworth HS9002A) that are modulated on and off using the AWG. Two channels of the microwave synthesizers are dedicated to producing the experiment's LO and a pulse at half the qubit's blue sideband transition frequency $\omega_{\textrm{sb}}/2$. The remaining two channels are dedicated to producing microwave pulses (300 ns in duration) at the qubit's ground to excited state transition frequency $\omega_{\textrm{ge}}$. The phase coherent microwave synthesizers are necessary for generating coherent propagating microwave fields (emitted from the cQED system) that encode the state of the qubit. However, these microwave synthesizers have frequency accuracy errors (at the 1~mHz level) that lead to phase drifts relative to the Agilent PSG synthesizers. Additionally, timing errors between the synthesizers and the AWG lead to phase errors. To reduce timing errors, all synthesizers are set to frequencies that are integer multiples of the protocol repetition rate (500 Hz). For constant phase drifts due to frequency accuracy errors, we separately measure and correct for such phase drifts after acquiring a set of measurements.

\begin{figure}[!ht]
  \centering
    \includegraphics[scale=0.75]{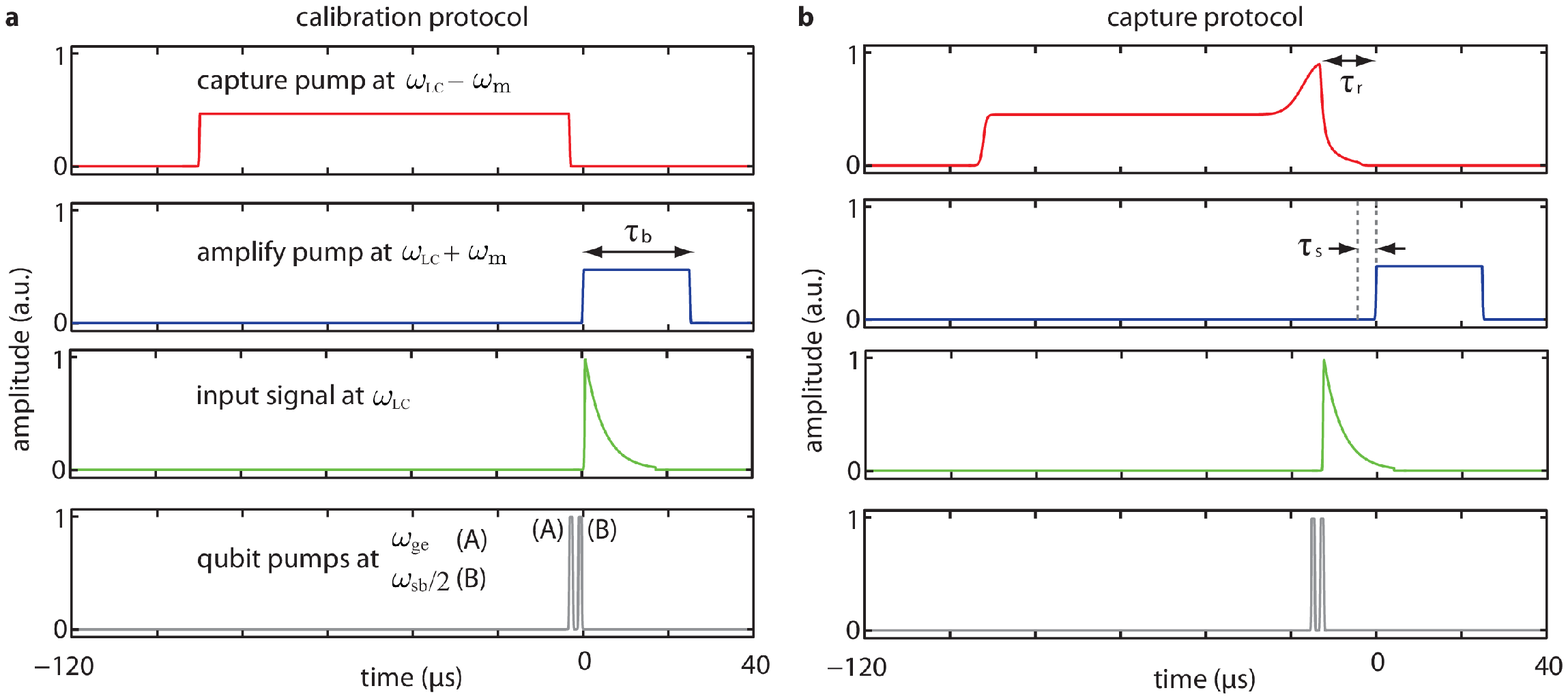}
      \caption{Temporal envelopes of the microwave pumps and signals. \textbf{a,} The timing diagram depicts the calibration protocol. Initially, a red-detuned `capture' pump at $\omega_{\textrm{LC}} - \omega_{\textrm{m}}$ cools the motion of the mechanical resonator. After the capture pump turns off, a blue-detuned `amplify' pump at $\omega_{\textrm{LC}} + \omega_{\textrm{m}}$ is pulsed on for a duration $\tau_{\textrm{b}}$. Input signals are coincident with the start of the amplify pump. \textbf{b,} The timing diagram for the capture protocol. For this protocol, the red-detuned pump is modulated for optimal capture of a signal with a decaying temporal envelope (as prescribed by Eqn. \ref{eqn:decaycatch}). The input signals (either generated by the microwave synthesizer or the cQED system) are coincident with the start of the modulation. The programmable storage time~$\tau_{\textrm{s}}$ is set by the delay between the end of the capture and the start of the amplification pump.}
\label{fig:TimingDiagram}
\end{figure}

\subsubsection{Detector}

Microwave signals incident or reflected off the electromechanical device are measured using a sensitive microwave detector. The detector consists of a Josephson parametric amplifier (JPA), high electron mobility transistor (HEMT) amplifier, a low noise room temperature amplifier (Miteq with +30 dB gain), a downconverting mixer, and a digitizer. For the protocols presented in Figs.~2 and 3 of the main text, we use the electromechanical device as the low-noise preamplifier instead of the JPA. The downconverting mixer is driven by a local oscillator detuned by 1 MHz from $\omega_{\textrm{LC}}$, and the inphase and quadrature channels of this mixer are sampled using a high speed digitizer (AlazarTech ATS 9462).

\subsection{Transmon qubit and microwave cavity parameters}

Embedding the transmon qubit inside a microwave cavity forms a circuit quantum electrodynamcis (cQED) system. We control and probe the cQED system by injecting microwave power into the weakly coupled port of the cavity, which couples to a transmission line at a rate $\kappa_{1}$. Microwave fields are emitted from the strongly coupled port, which couples to a separate transmission line at a rate $\kappa_{2}$. For our cQED system, the detuning between the cavity frequency ($\omega_{\textrm{c}}$) and the qubit frequency ($\omega_{\textrm{q}}$) is greater than their mutual coupling. In this regime, the cQED system is described by the dispersive Hamiltonian, $H_{\textrm{d}} = \omega_{\textrm{c}} a^{\dagger} a+ \omega_{\textrm{q}} \sigma_z /2 + \chi a^{\dagger} a \sigma_z$ where $\chi$ is the dispersive shift, $a$ describes the cavity mode, and $\sigma_z$ is a Pauli operator\cite{Blais2004}. The parameters defined by $H_{\textrm{d}}$, as well as the cavity parameters, were measured by performing spectroscopy on both the transmon qubit and the cavity. Single-shot measurements in the time-domain were used to measure the transmon qubit's energy relaxation time $T_1$ and the coherence time $T_{2}^\star$. The parameters for the cQED system are given in Supplementary Table \ref{tab:qubitcavparams}.

\begin{table}
\caption{\label{parameters}Parameters of the cQED system.}
\begin{ruledtabular}
\begin{tabular}{c  l  l }
{\bf{Symbol}}	&	{\bf{Description}}	&	{\bf{Value and units}} \\
$\omega_{\mathrm{c}}/2\pi$	&	Cavity resonant frequency (high power)	&	7.276\,781 GHz	\\
$\omega_{\mathrm{g}}/2\pi$	&	Cavity resonant frequency (qubit in ground state)	&	7.290\,156 GHz	\\	
$\omega_{\mathrm{e}}/2\pi$	&	Cavity resonant frequency (qubit in excited state)	&	7.283\,360 GHz	\\
$\kappa_{\mathrm{1}}/2\pi$	&	Weakly coupled cavity port decay rate	&	$<200$ Hz	\\	
$\kappa_{\mathrm{2}}/2\pi$	&	Strongly coupled cavity port decay rate	&	$50$ kHz	\\	
$\kappa_{\mathrm{c}}/2\pi$	&	Total cavity bandwith	&	60 kHz	\\	
$\chi/2\pi$	&	Dispersive shift	&	3.413 MHz	\\
$\omega_{\mathrm{eg}}/2\pi$	&	Ground to excited state transition frequency	&	5.652 MHz	\\	
$\omega_{\mathrm{sb}}/2\pi$	&	Blue sideband transition frequency	&	$2\times6.462$ GHz	\\	$K_\textrm{q}/2\pi$	&	Qubit anharmonicity	&	340 MHz	\\
$T_1$	&	Energy relaxation time	&	$\sim60$ $\mu$s	\\
$T_2^{\star}$	&	Coherence time	&	$\sim14$ $\mu$s	\\
\end{tabular}
\end{ruledtabular}
\label{tab:qubitcavparams}
\end{table}

\subsection{Electromechanical circuit parameters}
\label{sec:electromechanicsParams}

Parameters of the LC circuit and the mechanical resonator are determined by studying the microwave response of the device. These parameters are presented in Supplementary Table \ref{tab:EMCparams}.

In the steady state, we probe the LC circuit using a vector network analyzer that provides a weak microwave tone. We inject the microwave tone into the LC circuit and sweep the frequency of the tone. From the reflected microwave signal, we obtain a response that is fit to a model containing the resonant frequency of the LC circuit ($\omega_{\textrm{LC}}$), the total linewidth ($\kappa_{\textrm{LC}}$), and the external coupling rate ($\kappa_{\textrm{ext}}$). 

We determine properties of the aluminium membrane by probing the microwave response of the electromechanical device. To measure the resonant frequency of the membrane, we inject a microwave pump into the LC circuit and monitor the resulting sidebands. When the aluminium membrane resonates at $\omega_{\textrm{m}}$, the modulation sidebands are above and below the pump frequency by $\omega_{\textrm{m}}$.  To measure the decay time of the membrane ($\kappa_{\textrm{m}}^{-1}$), we used a pulsed ringdown method as described in Ref. \cite{Lecocq2015a}. Finally, we can measure the electromechanical coupling rate ($g_0$) and the average occupancy of the drumhead ($n_{\textrm{m}}$) by varying the temperature of the membrane's thermal environment provided by the dilution refrigerator\cite{Teufel2011}. At temperatures below 20~mK, we find the equilibrium occupation of the drumhead is approximately 42 quanta (Supplementary~Fig.~\ref{fig:MechTempSweep}).

\subsection{Capture, storage, and amplification of microwave fields}
\label{sec:electromechanicsEOM}
We operate the electromechanical device as an amplifier in order to characterize the states of either the input microwave field or of the mechanical resonator. For the calibration protocol, we use the electromechanical device to amplify an input propagating microwave field. For the capture protocol, we convert the state of the input microwave field to mechanical motion. We then use the electromechanical device to simultaneously amplify both the mechanical resonator state and convert it back into a microwave field. This section provides a brief mathematical description of these two protocols.

We pump the electromechanical device using a strong microwave field and model the dynamics using a set of coupled linear differential equations\cite{Aspelmeyer2014}. These equations describe the position of the mechanical resonator $x(t) = x_{\textrm{zp}} \left(c(t) + c^{*}(t)\right)$ and small fluctuations of the LC circuit's field $b(t)$ about the strong pump. Additionally, the LC circuit couples to propagating microwave fields $b_{\mathrm{in}}(t)$ and $b_{\mathrm{out}}(t)$. In a frame rotating at the pump's frequency, the equations of motion are
\begin{align}
\label{eqn:electromechanicsEOM1}
\dot{b}(t)&=\left(i\Delta-\kappa_{\textrm{LC}}/2\right)b(t)-i g (c(t)+c^* (t))+\sqrt{\kappa_{\mathrm{ext}}}b_{\mathrm{in}}(t) \nonumber \\
\dot{c}(t)&=(-i\omega_{\mathrm{m}}-\kappa_{\mathrm{m}}/2)c(t)-i\left(g b^{*}(t)+g^* b(t)\right) \nonumber \\
b_{\mathrm{out}}(t)&=\sqrt{\kappa_{\mathrm{ext}}}b(t)-b_{\mathrm{in}}(t)
\end{align} 
where $\Delta$ is the difference between the LC circuit's center frequency and $g = g_0 \beta$ is the coupling between the LC circuit and mechanical resonator, where $g_0$ is the electromechanical coupling rate and $n(t) = |\beta(t)|^2$ is the number of excitations induced in the LC circuit by the pump. Parameters in Eqns.~\ref{eqn:electromechanicsEOM1} are provided in Supplementary Table \ref{tab:EMCparams}.

We create an interaction capable of amplifying the mechanical and microwave modes by fixing the pump's detuning to~\mbox{$\Delta=+\omega_{\textrm{m}}$}. Because we operate the electromechanical device in the resolved-sideband ($4 \omega_{\textrm{m}} \gg \kappa_{\textrm{LC}}$) and weak-coupling ($|g| \ll \kappa_{\textrm{LC}}$) regimes, we can make approximations to Eqns.~\ref{eqn:electromechanicsEOM1} and express these equations in terms of $\Gamma_{\textrm{b}}(t) = 4 g_0^2 n_{\textrm{b}}(t)/\kappa_{\textrm{LC}}$ where $n_{\textrm{b}}$ is the number of photons induced in the LC circuit by the blue-detuned pump. In particular, we make a rotating wave approximation (neglect terms oscillating at~$2 \omega_{\textrm{m}}$), eliminate the LC circuit dynamics using an adiabatic approximation (set $\dot{b}(t) = 0$), and neglect the dampening of the mechanical resonator ($\kappa_{\textrm{m}} \ll \Gamma_{\textrm{b}}$). After making these approximations and moving into a frame rotating at the LC circuit's resonant frequency, Eqns.~\ref{eqn:electromechanicsEOM1} reduce to
\begin{align}
\label{eqn:electromechanicsEOM2}
\dot{c}^{*}(t)&= {1 \over 2} \Gamma_{\textrm{b}}(t) c^{*}(t) + \sqrt{\eta \Gamma_{\textrm{b}}(t)} b_{\textrm{in}}(t)e^{-i \psi_{\textrm{b}}} \nonumber \\
b_{\textrm{out}}(t)&= \sqrt{\eta \Gamma_{\textrm{b}}(t)} e^{i \psi_{\textrm{b}}} c^{*}(t) + (2 \eta - 1) b_{\textrm{in}}(t)
\end{align}
where $\eta = \kappa_{\textrm{ext}}/\kappa_{\textrm{LC}}$ is the coupling efficiency and $\psi_{\textrm{b}} = \arg (-i \beta_{\textrm{b}} )$ is determined by phase of the blue-detuned pump.

Eqns.~\ref{eqn:electromechanicsEOM2} describe how the electromechanical device can amplify propagating microwave fields and mechanical motion. For the calibration and capture protocols (see Supplementary~Fig.~\ref{fig:TimingDiagram}), we use a pulsed blue-detuned pump of the form
\begin{equation}
\Gamma_{\textrm{b}}(t)=\Gamma_{\textrm{0}}\Theta(t) \Theta(\tau_{\textrm{b}}-t)
\label{eqn:bluepump}
\end{equation}
where $\Gamma_{\textrm{0}}$ is a constant coupling rate, $\tau_{\textrm{b}}>0$ is the duration of the pump, $\Theta$ is the Heaviside step function. Using this blue-detuned pump, and assuming the initial state of the mechanical mode is $c^{*}(0)$, Eqns.~\ref{eqn:electromechanicsEOM2} have a solution given by
\begin{equation}
b_{\textrm{out}}(t) = c^{*}(0) \sqrt{\eta \Gamma_{\textrm{0}}} \cdot e^{i \psi_{\textrm{b}}} h(t) + \Gamma_{\textrm{0}} \eta  \cdot (h \star b_{\textrm{in}})(t) + (2 \eta - 1) b_{\textrm{in}}(t)
\label{eqn:EOM2solution}
\end{equation}
where $h(t) = \exp \left(\Gamma_{\textrm{0}}t/2 \right)$ and $\star$ denotes the convolution operation. Eqn.~\ref{eqn:EOM2solution} shows that the output field depends on both the states of the resonator and the input field.

While executing the calibration protocol, $b_{\textrm{in}}(t)$ is coincident with the start of $\Gamma_{\textrm{b}}(t)$.  For our tests using coherent signals, we choose the temporal and spectral content of $b_{\textrm{in}}(t)$ to match the bandwidth and center frequency of the cQED system. In a rotating frame, these fields are of the form
\begin{equation}
b_{\textrm{in}}(t) = \sqrt{\gamma_{\textrm{}}} B e^{ -\gamma_{\textrm{}}t/2}\Theta(t)
\end{equation}
where
$
\int_{0}^{\infty}|b_{\textrm{in}}(t)|^2 \diff t = |B|^2
$
is the total energy of the propagating mode and $\gamma$ is its power decay rate. Prior to amplification, we do not capture the propagating mode and so $c^{*}(0) = 0$ (here we neglect any fluctuations in the mechanical resonator's motion).

The expression for the output field takes on a simple form when we choose $\gamma_{\textrm{}} = \Gamma_{\textrm{0}}$. In this case, Eqn.~\ref{eqn:EOM2solution} reduces to
\begin{equation}
b_{\textrm{out}}(t) = 2 \sqrt{\eta \Gamma_{\textrm{0}}} B \sinh \left( \Gamma_{\textrm{0}}t/2 \right) + (2\eta-1)b_{\textrm{in}}(t) 
\end{equation}
The above expression shows that the temporal envelope of the output field is approximately a rising exponential, and the phase of the output field is independent of the blue-detuned pump's phase.

Instead of directly amplifying an input microwave field, we can capture its state in the mechanical resonator and then produce an amplified output microwave field whose state depends on the captured mechanical state. In this case, we use the capture protocol. We first inject a microwave field, $b_{\textrm{in}}(t+\tau_{\textrm{r}})$, that is delayed in time by $\tau_{\textrm{r}}>0$ relative to the start of the amplification pump, $\Gamma_{\textrm{b}}(t)$. As described in Ref.~\cite{Palomaki2013a}, a red-detuned pump set at $\Delta = -\omega_{\textrm{m}}$ creates a coupling between the input microwave field and the mechanical resonator at a rate given by $\Gamma_{\textrm{r}}(t) = 4 g_0^2 n_{\textrm{r}}(t)/\kappa_{\textrm{LC}}$ where $n_{\textrm{r}}(t)$ is the strength of the red-detuned pump. To optimally capture a signal with a decaying temporal envelope\cite{Andrews2015}, we modulate $n_{\textrm{r}}(t)$ so that
\begin{equation}
\label{eqn:decaycatch}
\Gamma_{\textrm{r}}(t+\tau_{\textrm{r}})  = \frac{\gamma_{\textrm{}} e^{-\gamma_{\textrm{}} (t+\tau_{\textrm{r}})}} {1 - e^{-\gamma_{\textrm{}} (t+\tau_{\textrm{r}})} + \gamma_{\textrm{}}/ \Gamma_{\textrm{r}}(0)} \Theta(t+\tau_{\textrm{r}})
\end{equation}
where $\gamma_{\textrm{}}/2\pi= 60$~kHz, $\Gamma_{\textrm{r}}(0)/2\pi\approx 1$~MHz, and $\tau_{\textrm{r}} = 30 \, \mu$s (for the protocols depicted in Fig.~2 of the main text). Using the parameters listed in Supplementary Table~\ref{tab:EMCparams}, we solve Eqns.~\ref{eqn:electromechanicsEOM2} and find that the fraction of the input signal's energy reflected off the LC circuit is $4.4\%$. For the data presented in Fig.~2b of the main text, we measure the fraction of reflected energy is 4.7$\%$.

After capture, the state of the microwave field is stored in the mechanical resonator and then converted back into a propagating microwave field. During storage, $\Gamma_{\textrm{r}}(t) = \Gamma_{\textrm{b}}(t) = 0$. By turning on $\Gamma_{\textrm{b}}(t)$ at $t=0$, the state of the mechanical resonator, $c^{*}(0)$, is converted to a propagating microwave field. While the amplification pump is on, $b_{\textrm{in}}(t) = 0$ and the output propagating field is
\begin{equation}
b_{\textrm{out}}(t) = c^{*}(0) \sqrt{\Gamma_{\textrm{0}} \eta} \cdot e^{i \psi_{\textrm{b}}} h(t).
\end{equation}
This result demonstrates that the amplified output microwave field contains the state of the mechanical resonator. Additionally, the output field has a phase that is conjugated relative to the directly amplified field.

The gain of amplification interaction can be adjusted by varying $r = \Gamma_{\textrm{b}} \tau_{\textrm{b}}$, as demonstrated in Supplementary~Fig.~\ref{fig:gainVSsqueezing}. In this figure, we plot the total energy of amplified signals, $E_{\textrm{out}} = \int_{-\infty}^{\infty} \langle V(t) \rangle^2 \diff t$ when the pumps were on, normalized to the energy of the input signal, $E_{\textrm{in}} = \int_{-\infty}^{\infty} \langle V(t) \rangle^2 \diff t$ when the pumps were off, as $\Gamma_\textrm{b}$ was varied. When the pumps were off, $V_{\textrm{dc}} = 0$, and the LC circuit was tuned out of resonance with the input signals. The model prediction was generated by discretizing and numerically integrating Eqns.~\ref{eqn:electromechanicsEOM1} for the total input and output energies as the magnitude of $\Gamma_{\textrm{b}}$ was varied. For our model prediction, we used the parameters outlined in Supplementary Table~\ref{tab:EMCparams}. We find the behavior of the amplification process (for large amplitude coherent signals) agrees well with our model.

\begin{figure}[!ht]
  \centering
    \includegraphics[scale=0.8]{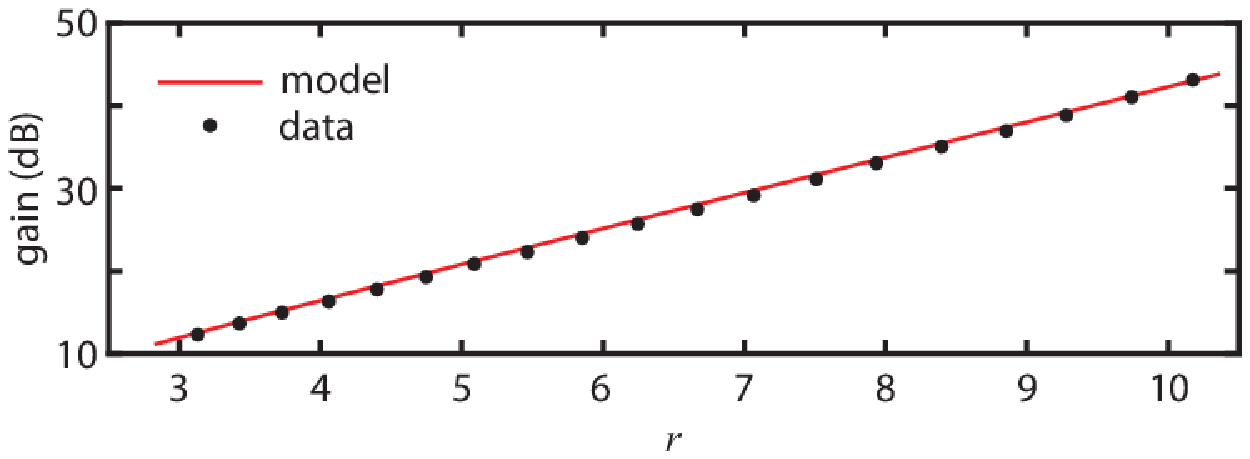}
      \caption{Adjustable gain of the mechanical amplification process. The plots shows the measured energy gain as $r = \Gamma_{\textrm{b}} \tau_{\textrm{b}}$ was varied.
      The measurements were acquired using the capture protocol. For these measurements, $\tau_{\textrm{b}} = 30$~$\mu$s was held constant while $\Gamma_{\textrm{b}}/2\pi$ was varied from 15~kHz to 55~kHz. The solid line indicates the expected performance obtained by numerically integrating Eqns.~\ref{eqn:electromechanicsEOM1} with all parameters determined separately.}
\label{fig:gainVSsqueezing}
\end{figure}

\begin{table}
\caption{Parameters of the electromechanical device. All parameters were measured at $V_{\mathrm{dc}}=5.83$~V.}
\begin{ruledtabular}
\begin{tabular}{c  l  l }
{\bf{Symbol}}	&	{\bf{Description}}	&	{\bf{Value and units}} \\
$\omega_{\mathrm{LC}}/2\pi$	&	Circuit resonant frequency	&	7.283\,360 GHz	\\
$\kappa_{\textrm{LC}}/2\pi$	&	Circuit decay rate	&	3 MHz	\\
$\kappa_{\mathrm{ext}}/2\pi$	 &	Circuit decay rate into the transmission line	& $2.59 \pm 0.01$ MHz		\\
$\omega_{\mathrm{m}}/2\pi$ & Mechanical resonant frequency &  9.345 MHz \\  
$\kappa_\mathrm{m}/2\pi$ & Mechanical decay rate & $14.5\pm1$ Hz  \\
$n_{\mathrm{m}}$& Average occupancy of the mechanical oscillator	& $42\pm2$	\\
$g_0/2\pi$	&	Electromechanical coupling	&	$283\pm14$ Hz \\			
\end{tabular}
\end{ruledtabular}
\label{tab:EMCparams}
\end{table}

\begin{figure}[!ht]
  \centering
    \includegraphics[scale=0.85]{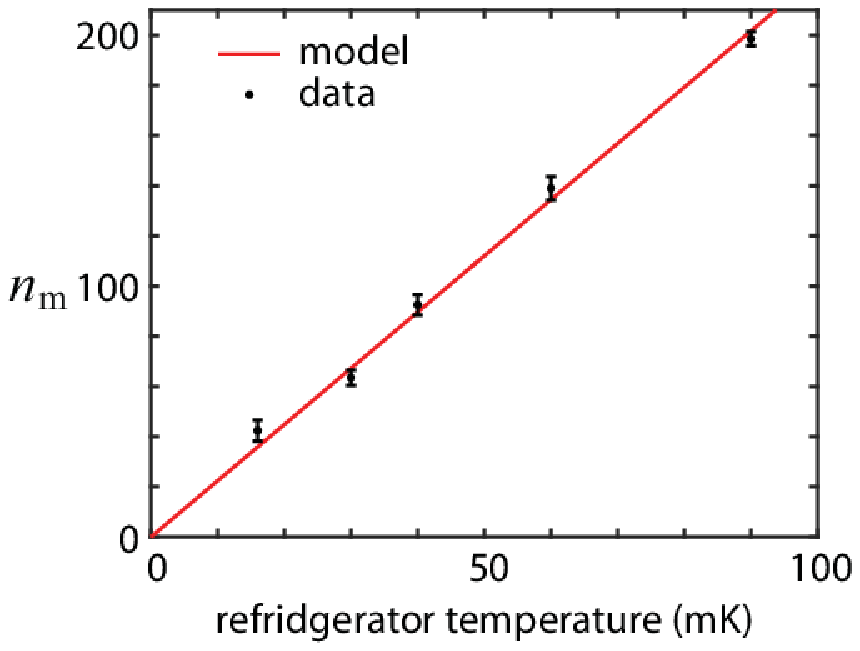}
      \caption{Electromechanical coupling rate and equilibrium occupation of the aluminium membrane. The black points show the average energy of the mechanical resonator (in units of quanta) as the refrigerator's base temperature was varied. Error bars show the standard error of the mean from five consecutive measurements. The solid line represents the expected mechanical occupation $n_{\textrm{m}}$ assuming the resonator is in equilibrium with its environment at temperatures between 30 mK and 100 mK and an electromechanical coupling rate $g_0/2\pi = 283\pm14$ Hz.}
      
\label{fig:MechTempSweep}
\end{figure}

\subsection{Quantum state estimation}

\subsubsection{Quadrature amplitude estimation}
\label{sec:ampEstimation}
Signals at the detector can be expressed as a pair of quadrature amplitudes $X$ and $Y$. During the amplification portion of either the calibration or capture protocol, the electromechanical device emits a signal that is routed to the detector. This signal has a known temporal envelope given by $f(t) = \exp \left( \Gamma_{\textrm{b}}t/2 \right)$ (see Supplementary~Section~\ref{sec:electromechanicsEOM}) and is mixed down to a frequency of $\omega_{\textrm{IF}}/2\pi = 1$ MHz. We digitally sample the mixed down signal at a rate of $R_s = 5$~MHz and form a discrete set of time-stamped voltages $\{t_k,V_k\}$ where $t_k = k\cdot T_s$ and $T_s = R_s^{-1}$ is the sampling period. By using our knowledge of the signal's spectral and temporal content, we define a pair of quadrature amplitudes\cite{Andrews2015}
\begin{align}
\label{eqn:quadamp}
X&=\sqrt{\frac{2 T_s}{\mathcal{G} C}}\sum\limits_{k=1}^{N} V_k f(t_k) \mathrm{cos}(\omega_{\textrm{IF}}t_k) \nonumber \\
Y&=\sqrt{\frac{2 T_s}{\mathcal{G} C}}\sum\limits_{k=1}^{N} V_k f(t_k) \mathrm{sin}(\omega_{\textrm{IF}}t_k)
\end{align}
where $C=\sum\limits_{k=1}^{N}  \left|f(t_k)\right|^2$, $N$ is the total number of samples, and $\mathcal{G}$ converts the voltage waveform into units of $\sqrt{\textrm{quanta} \cdot \textrm{Hz}}$. In the next section, we describe the procedure that we used to determine $\mathcal{G}$.

\subsubsection{Scaling the quadrature amplitudes}
\label{sec:scaleQuadAmps}
% The histogram labeled `no~photon' (Fig.~3 of the main text) corresponds to a reference measurement.
For each repetition of the calibration and capture protocols, we make a reference measurement in which vacuum fluctuations of the microwave field are injected into the electromechanical device. For this reference measurement, we do not generate single photons using the cQED system. Instead, we inject Johnson noise emitted from a 50~$\Omega$ load that is thermally anchored to the base stage of the dilution refrigerator held at a temperature $T$. For $T<25$~mK and near the frequency $\omega_{\textrm{LC}}$, the fluctuations in the microwave fields emitted from the load approach that of a vacuum state\cite{Clerk2010}. As such, we make the approximation that this input microwave mode is in an ideal vacuum state described by the density operator $\rho_{0}  = |0\rangle \langle0|$.

We use the reference measurements to determine $\mathcal{G}$, which relates voltage fluctuations at the output of the detector to vacuum fluctuations at the input of the electromechanical device. For the reference data set, we acquire a set of uncalibrated quadrature amplitudes that have a total voltage variance $\sigma_{\textrm{V}}^2$. If both the microwave and mechanical modes of the electromechanical device were in pure vacuum states, then we would expect the voltage fluctuations measured at the detector to have a total variance that corresponds to 1 quanta. For this case, we would calculate $\mathcal{G}$ so that $\mathcal{G}^{-1} \sigma_{\textrm{V}}^2~= 1$. However, even after cooling the mechanical resonator, we find that it is in a weak thermal state with an average occupation $n_{\textrm{th}}$. We take into account this estimate by calculating $\mathcal{G}$ so that $\mathcal{G}^{-1} \sigma_{\textrm{V}}^2~= 1 + n_{\textrm{th}}$. We then use $\mathcal{G}$ to scale the quadrature amplitudes (defined by Eqns.~\ref{eqn:quadamp}) for both data sets to have units of $(\textrm{quanta})^{1/2}$. To minimize systematic errors in the estimate of $\mathcal{G}$ due to potential drifts during the measurement, we alternate between the vacuum reference and single photon measurements every 512~executions of each protocol.

\begin{figure}[!ht]
  \centering
    \includegraphics[scale=0.9]{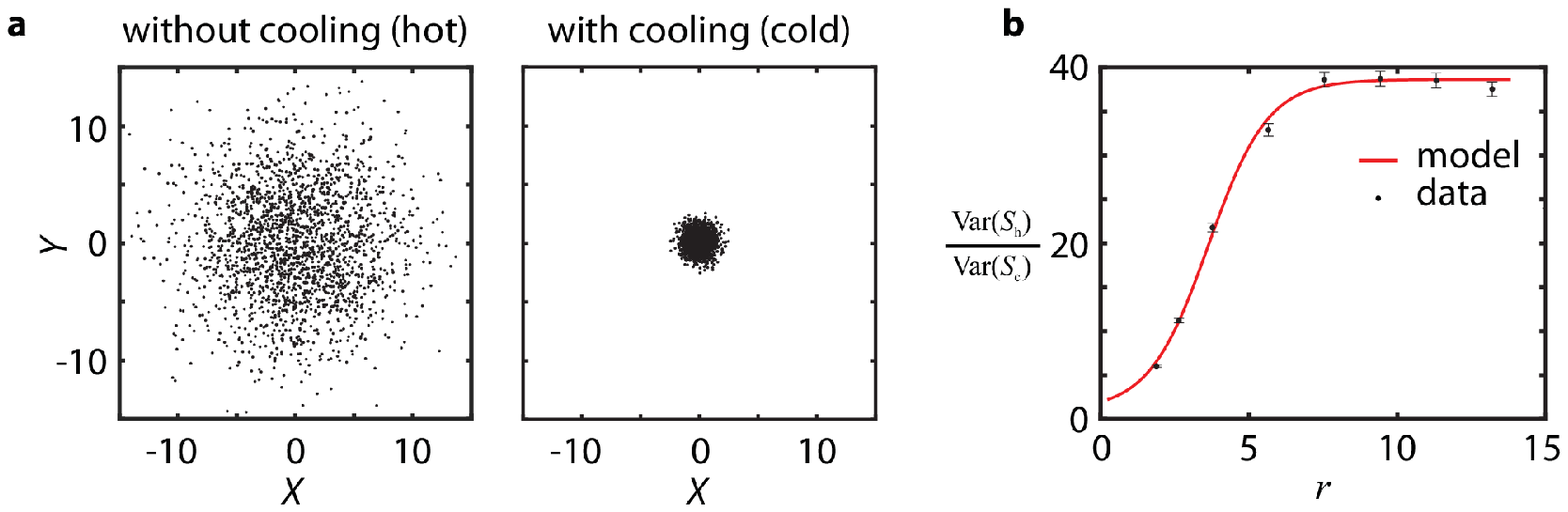}
      \caption{Amplified thermal states. \textbf{a,} The figure shows scatter plots of measured quadrature amplitudes $S = X + iY$ of signals at the detector. Without cooling the mechanical mode, the thermal motion of the mechanical resonator is amplified while the blue-detuned pump is pulsed on. The quadrature amplitudes ($S_{\textrm{h}}$, hot) are shown in the `without cooling' plot. Cooling the mechanical mode results in a reduced scatter in the quadrature amplitudes ($S_{\textrm{c}}$, cold) as shown in the `with cooling' plot. \textbf{b,} The plot shows the ratios of the total variance of the quadrature amplitudes without cooling, $\textrm{Var}(S_{\textrm{h}})$, and with cooling, $\textrm{Var}(S_{\textrm{c}})$, as $r=\Gamma_{\textrm{b}} \tau_{\textrm{b}}$ was varied. For these measurements, $\Gamma_{\textrm{b}}/2\pi = 60$~kHz while $\tau_{\textrm{b}}$ was varied from 5~$\mu$s to 55~$\mu$s. Once $r\gg1$, the plotted ratio approaches the expression given by Eqn.~\ref{eqn:hotcoldratio}. Error bars correspond to the statistical uncertainty in estimating the variance of 2,048 measurements.}  
\label{fig:HotColdVarPlots}
\end{figure}

We estimate $n_{\textrm{th}}$ by amplifying thermal states of two different temperatures. These thermal states describe the fluctuations of the mechanical mode during two types of measurements. In the first set of measurements, the mechanical mode is left in thermal equilibrium with its environment for over $30/ \kappa_{\textrm{m}}$. As shown in Supplementary Fig. \ref{fig:MechTempSweep}, we observe that the aluminium membrane thermalizes to an average mechanical occupation of $n_{\textrm{m}}=42$ quanta. We turn on the blue-detuned pump to create the amplification interaction, and then measure a set of quadrature amplitudes that have a total variance described by $\textrm{Var}(S_{\textrm{h}})$. In a second set of measurements, we cool the mechanical mode to nearly its quantum ground state by using the red-detuned pump. We again turn on the amplification interaction and obtain a total variance $\textrm{Var}(S_{\textrm{c}})$. For $r \gg 1$, the ratio of the variances approaches
\begin{equation}
{\textrm{Var}(S_{\textrm{h}}) \over \textrm{Var}(S_{\textrm{c}})} = {n_{\textrm{m}} + 1 \over n_{\textrm{th}} + 1}.
\label{eqn:hotcoldratio}
\end{equation}
If the mechanical mode is in its ground state, then $n_{\textrm{th}} = 0$ and Eqn.~\ref{eqn:hotcoldratio} yields $n_{\textrm{m}} + 1$. From our measurements, we find that this ratio is slightly reduced (Supplementary~Fig.~\ref{fig:HotColdVarPlots}). We attribute this reduction to a small residual occupation in the mechanical mode that is $n_{\textrm{th}} = 0.09\pm0.01$~quanta.

\subsubsection{Maximum likelihood method}

As explained in the previous sections, the data produced from an experimental run takes the form of a list of joint quadrature values $\{(X_k,Y_k):k=1,...,N\}$. The goal of quantum state tomography is then to convert the data into an estimate of the density matrix describing the state. For the calibration protocol, the state of the input microwave field is to be inferred, whereas for the capture protocol, the state of the mechanical resonator is to be inferred. These states are estimated via the method of Maximum Likelihood (ML).

In general, given a set of $N$ measurement outcomes $\{x_k\}$, each outcome is described mathematically by a positive-operator valued measure (POVM) element $E_k$ \cite{Nielsen2011}. If the system is in the state given by density matrix $\rho$, the probability of observing measurement outcome $x_k$ is $\mathrm{Pr}(x_k)=\mathrm{tr}(\rho E_k)$. The probability of obtaining the entire data set is then the product of the probabilities of each measurement outcome. This product defines the Likelihood function
\begin{equation}
\label{eqn:Likelihood}
L = \prod_k \mathrm{tr}(\rho E_k).
\end{equation}
The goal of ML tomography is to find the density matrix which maximizes the Likelihood function, or equivalently the log-Likelihood function, given by
\begin{equation}
\label{eqn:LogLikelihood}
\mathcal{L} = \sum_k \mathrm{log}(\mathrm{tr}(\rho E_k)).
\end{equation}
We maximize the log-Likelihood function with the R$\rho$R algorithm\cite{Hradil1997}. The algorithm begins with the maximally mixed state $\rho^{(0)} = \mathbb{I}/d$ as the initial estimate, where $\mathbb{I}$ is the identity matrix and $d$ is the Hilbert space dimension. With each iteration, $\rho$ is then updated according to 
\begin{equation}
\rho^{(i+1)} = \mathcal{N}\,\mathrm{R}(\rho^{(i)})\rho^{(i)}\mathrm{R}(\rho^{(i)}),
\end{equation}
where $\mathcal{N}$ is an overall normalization to ensure that $\mathrm{tr}(\rho)=1$ at each step, and $\mathrm{R}(\rho)$ is a matrix given by
\begin{equation}
\mathrm{R}(\rho)=\frac{1}{N}\sum_k \frac{E_k}{\mathrm{tr}(\rho E_k)}.
\end{equation}
The R$\rho$R transformation leaves the maximum-likely state unchanged, and while the algorithm is not guaranteed to converge\cite{Jaroslav2007}, it does so in almost all practical cases, including the case considered here.

In our experiment, the set of joint quadratures values $\{(X_k,Y_k)\}$ constitutes a set of simultaneous `position' and `momentum' measurements.  In the optical domain, such a measurement would be performed by splitting a state on a beamsplitter and measuring a single quadrature of each output mode\cite{Leonhardt1997}. The microwave equivalent of this technique is phase-insensitive linear amplification\cite{Eichler2012}. In the capture protocol, the amplification process is nearly quantum-limited, because the microwave mode is nearly in a vacuum state. In this case, the probability density for obtaining a pair of quadrature values $(X_k,Y_k)$ is given by the Husimi Q-function\cite{Leonhardt1993}: 
\begin{equation}
\mathrm{Pr}(X_k,Y_k)=Q(X_k,Y_k)=\frac{1}{\pi}\bra{\alpha_k}\rho\ket{\alpha_k}=\frac{1}{\pi}\mathrm{tr}(\rho\ket{\alpha_k}\bra{\alpha_k})
\end{equation}
Here $\ket{\alpha_k}$ is a coherent state with $\alpha_k=X_k+iY_k.$ From the above equation, note that the POVM elements for the measurement outcomes are projections onto coherent states:
\begin{equation}
E_k=\frac{1}{\pi}\ket{\alpha_k}\bra{\alpha_k}.
\end{equation}
For the calibration protocol, we use the electromechanical device to directly amplify the input microwave field.  When operated in this manner, we assume the added noise of the amplifier is a result of the occupancy of the mechanical mode. We approximate this occupancy to be $n_\mathrm{th}=0.1$ quanta (see Supplementary~Section~\ref{sec:scaleQuadAmps}). In this case, the POVM operators are projections onto displaced thermal states, rather than projections onto displaced vacuum (coherent) states:
\begin{equation}
\label{eqn:noisyPOVM}
E_k=\frac{1}{\pi}D(\alpha_k)\rho_{\mathrm{th}}D^{\dagger}(\alpha_k).
\end{equation}
where $D(\alpha) = \exp \left( \alpha a^{\dagger}-\alpha^* a \right)$ is the displacement operator and $\rho_{\mathrm{th}}$ is a thermal state with thermal occupancy $n_\mathrm{th}$. To derive Eqn.~\ref{eqn:noisyPOVM}, we note that because the added noise of the amplifier is not quantum limited, the probability density for obtaining quadrature values $\{(X_k,Y_k)\}$ is no longer the Q-function of the input microwave mode but rather the convolution of the Q-function with the added thermal noise:
\begin{align}
        \mathrm{Pr}(X,Y)&=\int d\alpha^{\prime}Q(\alpha^{\prime})e^{-\abs{\alpha-\alpha^{\prime}}^2/2n_\mathrm{th}}\nonumber \\
        &= \int d\alpha^{\prime} \frac{1}{\pi}\mathrm{tr}(\rho\ket{\alpha^{\prime}}\bra{\alpha^{\prime}})e^{-\abs{\alpha-\alpha^{\prime}}^2/2n_\mathrm{th}} \nonumber \\
        &= \frac{1}{\pi}\mathrm{tr}\int d\alpha^{\prime}\rho\ket{\alpha+\alpha^{\prime}}\bra{\alpha+\alpha^{\prime}}e^{-\abs{\alpha^{\prime}}^2/2n_\mathrm{th}} \nonumber \\
        &= \frac{1}{\pi}\mathrm{tr}\rho D(\alpha)\bigg(\int d\alpha^{\prime}\ket{\alpha^{\prime}}\bra{\alpha^{\prime}}e^{-\abs{\alpha^{\prime}}^2/2n_\mathrm{th}}\bigg)D^{\dagger}(\alpha) \label{eqn:noisyPOVMderive}
\end{align}
The third equality follows from a change of variables and the integral in the fourth line of Eqn. \ref{eqn:noisyPOVMderive} is equivalent to a thermal state $\rho_{\mathrm{th}}$ with mean excitation number $n_{\mathrm{th}}$ as discussed in Ref.~\cite{Leonhardt1997}.

In our execution of the R$\rho$R algorithm, we truncate the Hilbert state space to dimension $d=16$ for both the calibration and capture data sets. We set this cutoff to be high enough for the maximum-likely density matrix elements to no longer depend on Hilbert space truncation. We run the algorithm for 500 iterations. We chose this number by generating synthetic data from various known density matrices and running ML with different numbers of iterations. We found that 500 iterations was sufficient for the density matrix elements to differ by no more than $10^{-3}$ from the density matrix elements obtained using significantly larger ($\sim$10,000) number of iterations.

\subsubsection{$g^{(2)}$ calculation}
\label{gTwoSec}

The density matrix elements presented in Fig.~3 of the main text can be used to calculate the degree of second-order coherence at zero time delay, $g^{(2)}$. This quantity is often used to characterize the statistical properties of photons emitted from a source\cite{Gerry2004}. For example, an ideal single photon source has $\rho_{11} = 1$ with all other elements equal to zero. This state yields $g^{(2)}(0) = 0$. In general,
\begin{equation}
g^{(2)}(0) = {\langle \hat{n}^2 \rangle - \langle \hat{n} \rangle \over \langle \hat{n} \rangle^2}
\label{eqn:g2number}
\end{equation}
where $\hat{n}$ is the number operator. In terms of the density matrix elements, Eqn.~\ref{eqn:g2number} becomes
\begin{equation}
g^{(2)}(0) = {\sum\limits_{n} n(n-1) \rho_{nn} \over \left( \sum\limits_{n} n \rho_{nn} \right)^2}
\label{eqn:g2rho}
\end{equation}
where $\rho_{nn} = \langle n|\rho |n\rangle$.

We use Eqn.~\ref{eqn:g2rho} with the diagonal density matrix elements obtained via tomography to calculate $g_{\textrm{m}}^{(2)} = 0.89$. We use the bootstrap error analysis described in Supplementary Section \ref{errorAnalysis} to obtain a 90\% confidence interval of [0.72,~0.94]. A histogram of bootstrapped values of $g_{\textrm{m}}^{(2)}$ is displayed in Supplementary~Fig.~\ref{fig:histsgTwo}. The density matrix obtained via tomography acts on a truncated Hilbert space with maximum Fock number $n=15$.  While the first 3 diagonal elements are presented in Fig. 3 of the main text, all 16 elements are used in the calculation of $g_{\textrm{m}}^{(2)}$. These elements rapidly become smaller with increasing $n$ such that both $g_{\textrm{m}}^{(2)}$ and the bounds of the confidence interval are independent of the Hilbert space truncation. We find that they converge to within 1\% of our reported values once $n>8$.

\begin{figure}[!ht]
  \centering
    \includegraphics[scale=1]{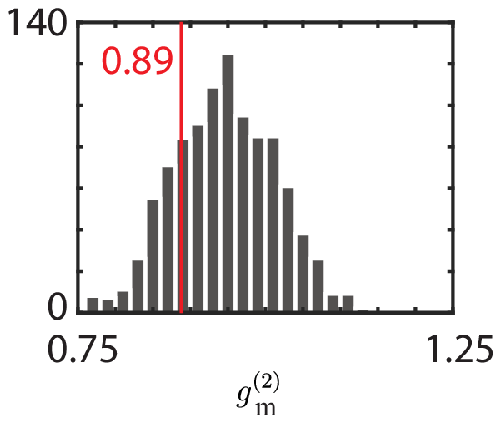}
      \caption{Bootstrapped histograms of the calculated $g^{(2)}$ function. The figure shows a histogram of $g^{(2)}_{\textrm{m}}$ for the captured mechanical state, obtained via bootstrapping on the density matrices estimated by ML tomography. The red bar indicates the value calculated from the estimated density matrices. The histogram is obtained by performing ML on 1,000 synthetic data sets with 102,400 measurements each. See Supplementary Section~\ref{errorAnalysis}.}
      
\label{fig:histsgTwo}
\end{figure}

\subsubsection{Error analysis}
\label{errorAnalysis}

We use a parametric bootstrap method to analyze the statistical error in the density matrix elements estimated by the ML tomography. The estimated density matrix $\rho_{\mathrm{est}}$ is used to generate synthetic data sets, and ML tomography is performed on each of the synthetic data sets, building a histogram of values for each density matrix element. Example histograms showing the bootstrapped diagonal density matrix elements for the mixed single photon state in the calibration and capture protocols are shown in Supplementary~Fig.~\ref{fig:bootstrap}. These histograms reveal an asymmetry in the bootstrapped distributions, as well as bias in some of the density matrix element estimates. It is therefore more appropriate to analyze the statistical error in terms of confidence intervals rather than the standard error. The reported error bars indicate 90\% \textit{basic bootstrap} confidence intervals obtained from these histograms \cite{Efron1993}. The basic bootstrap confidence interval is obtained as follows: let $\theta$ be the estimated parameter and let $\theta_{\mathrm{lo}}$ and $\theta_{\mathrm{up}}$ be the lower and upper percentile values obtained from the bootstrapped histogram. That is, $\theta_{\mathrm{lo}}$ ($\theta_{\mathrm{up}}$) is the value for which $5\%$ of bootstrapped values are less than (greater than) $\theta_{\mathrm{lo}}$ ($\theta_{\mathrm{up}}$). The differences $\alpha = \theta-\theta_{\mathrm{lo}}$ and $\beta = \theta_{\mathrm{up}} - \theta$ between the estimated parameter value and the lower and upper percentile values are then inverted around the estimated parameter $\theta$ to obtain the confidence interval $[\theta-\beta,\,\theta+\alpha]$.  

For each inferred density matrix we create 1,000 synthetic data sets, with each data set containing 102,400 simulated measurement outcomes. The ML tomography algorithm is applied to each synthetic data set with the same number of iterations and Hilbert space truncation as was used on the experimental data generated by executing either the calibration or capture protocols.

To generate synthetic data sets, we use a Monte Carlo method to sample pairs of joint quadrature values $(X_k,Y_k)$ from the Husimi Q-function $Q(X_k,Y_k)$ corresponding to the estimated density matrix $\rho_{\mathrm{est}}$. A uniformly distributed set of random points $(X_k,Y_k)$ all lying within a sufficiently large radius from the origin in phase space is generated. The Q-function is computed for each of these points, and points are then discarded with probability $$1-\frac{Q(X,Y)}{ \max
\limits_k  Q(X_k,Y_k)}$$ so that the remaining points are distributed according to the Q-function. For the data sets obtained from the calibration protocol, the additional thermal noise $n_{\mathrm{th}}$ in the mechanical mode must be taken into account. In this case, the measured joint quadrature values are sampled not from the Q-function of the microwave mode, but from the Q-function convolved with the added thermal noise in the mechanical mode. We account for this added noise in our Monte Carlo sampling method by simply adding Gaussian noise to each sample.

Our finite precision in calibrating the thermal noise in the mechanical mode (Supplementary~Fig.~\ref{fig:HotColdVarPlots}) is a source of systematic error in the tomographic estimate of density matrix elements. Both the procedure for rescaling histograms as well as the tomography assume a mechanical thermal occupancy of $n_{\mathrm{th}}=0.1$ quanta. We therefore investigate how the density elements would change if the value of $n_{\mathrm{th}}$ were different. We let $n_{\mathrm{th}}$ range from 0.08 to 0.12 in steps of 0.01, and for each step we obtain the density matrix for both the input electrical and converted mechanical states. The diagonal density matrices obtained with different values for $n_\mathrm{th}$ are shown in (Supplementary~Table~\ref{tab:errorsNthRho}). We find that these density matrix elements change linearly and by not more than $\sim0.02$ over the range in $n_{\mathrm{th}}$ that we explore.

\begin{figure}[!ht]
  \centering
    \includegraphics[scale=0.9]{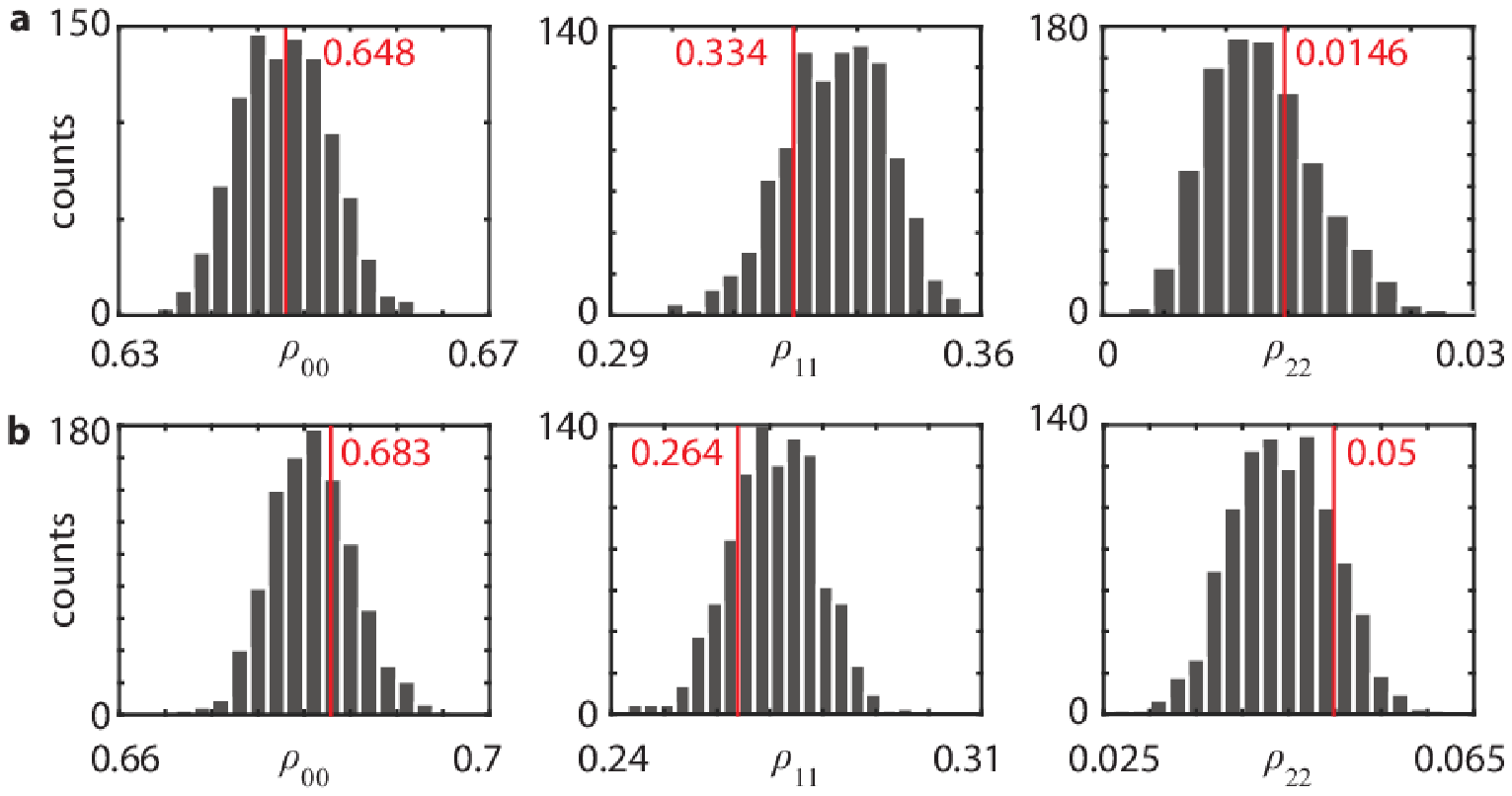}
      \caption{Bootstrapped histograms.
      \textbf{a,} This figure shows the histograms of the $\rho_{00}$, $\rho_{11}$, and $\rho_{22}$ density matrix elements obtained via bootstrapping on the estimated state of the input microwave field in the mixed single photon data set. The red bars indicate the values obtained by ML from the experimental data. Each histogram is obtained by performing ML on 1,000 synthetic data sets with 102,400 measurements each. \textbf{b,} This figure shows similar histograms, but for the mechanical resonator state.}
      
\label{fig:bootstrap}
\end{figure}

\begin{table}
\caption{Estimated density matrix elements assuming different values of $n_{\textrm{th}}$ in the quadrature amplitude scaling and tomography procedures (see Supplementary~Section~\ref{errorAnalysis}).}
\begin{ruledtabular}
\begin{tabular}{c  l  l  l  l  l  l}
$n_{\textrm{th}}$	&	$\left[ \rho_{\textrm{e}} \right]_{00}$	&	$\left[ \rho_{\textrm{e}} \right]_{11}$ & $\left[ \rho_{\textrm{e}} \right]_{22}$ & $\left[ \rho_{\textrm{m}} \right]_{00}$ & $\left[ \rho_{\textrm{m}} \right]_{11}$ & $\left[ \rho_{\textrm{m}} \right]_{22}$
\vspace{2mm} \\
0.08 & 0.655 & 0.328  & 0.014  & 0.694 & 0.266 & 0.038\\
0.09 & 0.652 & 0.331 & 0.014 & 0.689 & 0.265 & 0.044\\
0.10 & 0.649 & 0.334  & 0.015  & 0.683 & 0.264 & 0.050\\
0.11 & 0.645 & 0.337 & 0.015 & 0.677 & 0.264 & 0.056\\
0.12 & 0.642 & 0.340  & 0.015  & 0.672 & 0.263 & 0.062\\
\end{tabular}
\end{ruledtabular}
\label{tab:errorsNthRho}
\end{table}

\subsubsection{Model of the capture process}
\label{beamsplitterModel}

The capture process described in section \ref{sec:electromechanicsEOM} creates a coupling between propagating microwave fields and mechanical motion. This process can be viewed as implementing a beamsplitter interaction between the microwave and the mechanical modes\cite{Palomaki2013a}.  A rudimentary model of the capture process therefore involves sending the state of the microwave field through one port of a beamsplitter and the state of the mechanical mode through the other input port. The transmissivity of the beamsplitter models imperfect capture of the propagating mode. For an ideal capture of the propagating microwave mode, there would be no reflected microwave power off of the electromechanical device. 
After tracing over the output microwave mode, the mechanical resonator would contain the state of the input propagating microwave mode.

A limitation of this model is that it does not take into account the non-zero occupancy of the mechanical mode. Ideally, an input vacuum state of the microwave field to a beamsplitter with perfect efficiency would result in a vacuum state of the mechanical mode. This model would describe the ideal cooling of the mechanical mode to its quantum ground state. However, we observe a small residual occupation in the mechanical mode $n_{\textrm{th}}$. A possible source of this residual occupation is the internal loss of the LC circuit due to a surface layer of two-level system fluctuators\cite{Gao2007}.  This internal loss couples the microwave mode to a thermal bath with a finite occupation.

We incorporate imperfect capture efficiency and internal loss of the LC circuit into a heuristic model, as shown in Supplementary Fig.~\ref{fig:model}. This model contains two beamsplitters, $B_1$ and $B_2$, which model the internal loss of the LC circuit and the capture efficiency, respectively. The input electrical state $\rho_{\mathrm{e}}$ and an ancillary electrical state $\rho_{\mathrm{an}}$ are sent through a beamsplitter $B_1$, and the reduced state of the output electrical mode is then mixed with a thermal state of the mechanical mode $\rho_{\mathrm{th}}$ on beamsplitter $B_2$. Explicitly, this complete process is $\rho_{\mathrm{out}}=\mathrm{tr}_{\mathrm{e}}(B_2\,\rho^{\prime}\otimes\rho_{\mathrm{th}} \,B_2^{\dagger})$ where $\rho^{\prime}=\mathrm{tr}_\mathrm{an}(B_1\,\rho_{\mathrm{an}}\otimes\rho_{\mathrm{e}} \,B_1^{\dagger})$. The ancilla state $\rho_\mathrm{an}$ is taken to be a thermal state whose thermal occupancy is determined implicitly by the requirement that a vacuum input state $\rho_{\textrm{e}}=\ket{0}\bra{0}$ yields a thermal intermediate state $\rho^{\prime}=\rho_\mathrm{th}$ with mean occupancy $n_\mathrm{th}=0.1$ quanta. In the Heisenberg picture, a beamsplitter transforms the annihilation operators for its two input modes $a_1$ and $a_2$ as $B^{\dagger}a_{1,2}B = a_{1,2}^{\prime}$, where
\begin{equation}
    \begin{pmatrix}
    a_1^{\prime}\\
    a_2^{\prime}
    \end{pmatrix}
    =
    \begin{pmatrix}
    \cos{\theta} & -\sin{\theta} \\
    \sin{\theta} & \cos{\theta}
    \end{pmatrix}
    \begin{pmatrix}
    a_1\\
    a_2
    \end{pmatrix}.
\end{equation}
Under this convention, $\theta=0$ gives the identity whereas $\theta=\pi/2$ corresponds to perfect capture. The beamsplitters in the model are thus each parametrized by an angle $\theta_i$ where $i=\{1,2\}$, which in turn is related to the \textit{reflection coefficient} $R_i=\mathrm{sin}^2\theta_i$.

For the data presented in Fig. 2b of the main text, we find that the fraction of energy reflected from the electromechanical device is approximately $5\%$. As such, we use $R_2 = 0.95$ for $B_2$. Note that a low amount of reflected energy corresponds to a large value of the beamsplitter reflection coefficient, because the reflection coefficient models the efficiency of mode swapping. As discussed in Supplementary Section \ref{sec:electromechanicsParams}, the fractional resonator energy loss is measured to be approximately 0.14 and so we use $R_1 = 0.14$ for $B_1$.

We use this cascaded beamsplitter model of the capture process to predict the captured mechanical states from the known input states obtained using the calibration protocol. For example, using the mixed single photon input state described in the main text, the model predicts $\left[ \rho_{\textrm{m}} \right]_{00} = 0.67$, $\left[ \rho_{\textrm{m}} \right]_{11} = 0.27$, and $\left[ \rho_{\textrm{m}} \right]_{22} = 0.05$, whereas the 90\% confidence intervals for these quantities obtained from tomography and bootstrapping are [0.68,~0.69], [0.24,~0.27], and [0.05,~0.06], respectively. We also use the model to estimate the bias in the average fidelity calculation. This bias is due to truncating the input and output density matrices to dimension $d=2$ (see~Supplementary~Section~\ref{averageFidelity}).     

We emphasize that our mathematical description of the capture process should be viewed as a heuristic model. This model relies on two free parameters, the LC circuit loss and capture efficiency, which were measured independently and may vary for each experimental run.  Nevertheless, to the degree that this model accurately describes the process, it provides prospects for improving the capture process fidelity. We expect that reducing the LC circuit loss, increasing the capture efficiency, and minimizing the occupation of the mechanical mode should yield a capture process that approaches unit fidelity.
% capture process model
\begin{figure}[!ht]
  \centering
    \includegraphics[scale=0.90]{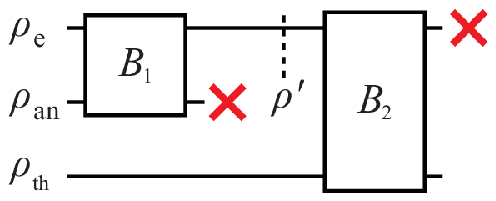}
    \caption{Model of the capture process. The diagram depicts the three modes and the two beamsplitter interactions used to model the capture process. The input state of the electrical mode is $\rho_{\mathrm{e}}$ and the ancilla state $\rho_{\mathrm{an}}$ describes the state of the additional electrical mode. The mechanical mode is in the thermal state $\rho_{\mathrm{th}}$. The electrical modes couple via a beamsplitter interaction $B_1$ and the states $\rho^{\prime}$ and $\rho_{\mathrm{th}}$ couple via another beamsplitter interaction $B_2$. A red cross denotes a partial trace.}
\label{fig:model}
\end{figure}

\subsubsection{Model of mechanically stored states}

We model the evolution of mechanically stored states by using a master equation formalism\cite{Gardner2004}. Using this formalism, we model the state of the mechanical resonator $\rho_{\textrm{m}}$ as a single damped harmonic oscillator coupled at a rate $\kappa$ to a thermal bath with an average occupation $N$. In the number basis, this model becomes a set of coupled ordinary differential equations given by
\begin{equation}
    {d P_{n} \over dt} = -\kappa\left[ (N + 1) n P_n(t) + N (n + 1) P_n(t) - (N + 1)(n + 1) P_{n+1}(t) - N n P_{n-1}(t)\right],
\label{eqn:mastereqn}
\end{equation}
where $P_n = \langle n | \rho_m | n\rangle$ describes the probability of observing $n$ excitations in the mechanical mode. For our system, we take $N = n_\textrm{m}$ and make the approximation $N + 1 \approx N$. This approximation allows us to parametrize Eqn. \ref{eqn:mastereqn} using the decoherence rate, $\gamma = \kappa N$, as a single parameter.

We use Eqn.~\ref{eqn:mastereqn} to model the evolution of mixed single phonon states stored in the mechanical mode, as shown in Fig. 3f of the main text. By discretizing and numerically integrating Eqn. \ref{eqn:mastereqn}, we obtain $P_n(t)$. In our model, the mechanical decoherence rate $\gamma_{\textrm{m}}$ is a free parameter and the estimated density matrix elements of the initial state are used to fix $P_{n}(0)$. Simultaneous fits to the inferred diagonal elements presented in Fig. 3f of the main text yields a characteristic storage time for a mixed photon state of $\tau_{\textrm{m}} = \gamma_{\textrm{m}}^{-1} = 137\pm6$~$\mu$s.

As a control experiment, we do not capture and store single photons. For this experiment, we execute the protocol depicted in Fig.~3d of the main text but without producing single photons using the cQED system. The results of this experiment and fits to the model are shown in Supplementary Fig.~\ref{fig:thermalStateStorage}. For the model, we assume the mechanical resonator is initially in a thermal state and remains thermal but with increasing average occupation $$\displaystyle \langle n \rangle = \sum_n n P_n$$ as it equilibrates with the thermal bath. In this case, the model takes on a simple form after using
\begin{equation}
    \frac{d}{dt} \langle n(t) \rangle =\sum_n n {d P_{n}(t) \over dt}
\label{eqn:averagen}
\end{equation}
and Eqn.~\ref{eqn:mastereqn} to obtain
\begin{equation}
    \frac{d}{dt} \langle n(t) \rangle =-\kappa \left(\langle n(t) \rangle-N\right).
\label{eqn:rateequation}
\end{equation}
Using the solutions to this rate equation, and the thermal distribution
\begin{equation}
    P_{n} = { \langle n \rangle^n \over \left(\langle n \rangle + 1\right)^{n+1}},
\end{equation}
we simultaneously fit the estimated density matrix elements shown in Supplementary~Fig.~\ref{fig:thermalStateStorage} and extract $\tau_{\textrm{m}} = 137\pm8$~$\mu$s.

\begin{figure}[!ht]
  \centering
    \includegraphics[scale=0.75]{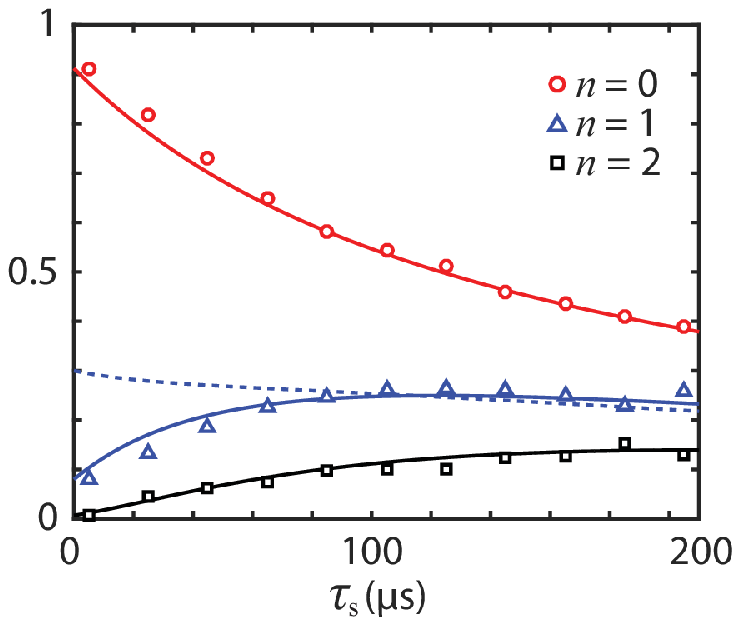}
      \caption{Evolution of a mechanical thermal state. The plot shows the first three diagonal density matrix elements of the extracted mechanical state $\rho_{\textrm{m}}$ as a function of the storage time $\tau_{\textrm{s}}$. The plot is similar to the one in Fig.~3g of the main text, but for the control experiment in which no single photons were generated and captured.  The solid lines indicate a model of a thermal state that uses $\tau_{\textrm{m}}$ as a free parameter. From the fit, we extract $\tau_{\textrm{m}} = 137\pm8$~$\mu$s. For comparison, the dashed line indicates the evolution of $P_1$ for the initial mixed phonon state presented in Fig.~3g of the main text.}
      
\label{fig:thermalStateStorage}
\end{figure}

\subsection{Quantum process fidelity estimation}

\subsubsection{Average fidelity}
\label{averageFidelity}

Any quantum process is described mathematically by a completely-positive trace preserving (CPTP) map $$\mathcal{E}:\rho\to\mathcal{E}(\rho).$$ The \textit{average fidelity} of a process is a measure of how well that process preserves quantum information. It is defined as
% average fidelity
\begin{equation}
F_{\mathrm{avg}}=\int d\Psi\bra{\Psi}\mathcal{E}(\ket{\Psi}\bra{\Psi})\ket{\Psi}.
\end{equation}
That is, the average fidelity gives the average overlap of the output of the process with the initial state, averaged over all pure states. A related quantity is the \textit{entanglement fidelity}, which is defined as follows: let $\ket{\phi}=\frac{1}{d^2}\sum\limits_i\ket{i}\ket{i}$ denote a maximally entangled state. Here, $\{\ket{i}\}$ is an orthonormal basis and $d$ is the dimension of the Hilbert space on which the process $\mathcal{E}$ acts. Then the \textit{entanglement fidelity} is
% entanglement fidelity
\begin{equation}
F_{\textrm{e}}=\bra{\phi}(\mathcal{E}\otimes \mathbb{I})(\ket{\phi}\bra{\phi})\ket{\phi}.
\end{equation}
The entanglement fidelity measures how well a system's entanglement with another system is preserved when the quantum process acts only on the first system. Explicit calculation shows that the entanglement fidelity can also be expressed as
\begin{equation}
\label{eqn:entanglement Fidelity}
F_{\textrm{e}}=\frac{1}{d^2}\sum\limits_{ij}\bra{i}\mathcal{E}(\ket{i}\bra{j})\ket{j}
\end{equation}
The utility of the entanglement fidelity is that it is easy to calculate, and there is a simple formula relating the average fidelity to the entanglement fidelity \cite{Nielsen2002}:
\begin{equation}
\label{eqn:Nielsen formula}
F_{\mathrm{avg}}=\frac{d\times F_{\textrm{e}}+1}{d+1}
\end{equation}
Because quantum processes act as linear operators on the space of operators, the fidelity can be determined if the outputs $\mathcal{E}(\rho_i)$ are known for each of a set of input states $\rho_i$ that span the space of operators on Hilbert space. For a Hilbert space of dimension $d$, $d^2$ linearly independent density matrices are required\cite{Nielsen2011}. To calculate the fidelity of the capture process in our experiment, we truncate the estimated input and output density matrices to a Hilbert space dimension of $d=2$. The input states are obtained by estimating the density matrix of the input microwave field after executing the calibration protocol. Similarly, the output states are obtained by estimating the density matrix of the mechanical resonator state after executing the capture protocol. We form a set of basis states by using the following input states: a vacuum state, a mixed single photon state, and mixed superposition states with phases chosen from the set $\{0,\pi/2\}$. We expand operators $\ket{i}\bra{j}$ in this basis. The average fidelity $F_\mathrm{avg}$ is then calculated using the known input and output states shown in Supplementary Table \ref{tab:densityMatrices} and Eqns. \ref{eqn:entanglement Fidelity}, and \ref{eqn:Nielsen formula}. These states are acquired in a separate experiment consisting of 20,480 measurements.

The formula for the average fidelity given by Eqn.~\ref{eqn:Nielsen formula} is actually only true if the process maps all states into a Hilbert space of dimension $d$, the same dimension as the domain of the process. In our experiment, the weak thermal occupation of the mechanical resonator leads to output states with small but non-zero density matrix elements for $n\geq2$, in which case truncating the output density matrices to dimension $d=2$ leads to a systematic overestimation of the average fidelity. We can account for this by using a more general expression $F_\mathrm{avg}$ which is valid in the case where the process takes states into a higher dimensional space than its domain:
\begin{equation}
\label{eqn:Average fidelity}
F_{\mathrm{avg}}=\frac{d}{d+1}(F_e+A(\mathcal{E}))
\end{equation}
where
\begin{equation}
\label{eqn:AofE}
A(\mathcal{E})=\frac{1}{d^2}\sum\limits_{ij}\bra{i}\mathcal{E}(\ket{j}\bra{j})\ket{i}.
\end{equation}
Eqn.~\ref{eqn:Average fidelity} can be motivated by observing that for a process which does not map states into a higher dimensional space, Eqn.~\ref{eqn:AofE} is a sum of traces of density matrices (which each have unit trace) and Eqn. \ref{eqn:Nielsen formula} is recovered. That Eqn.~\ref{eqn:Average fidelity} is correct for processes which map states out of their domain can be directly verified for simple processes such as the `erasure channel,' given by $\mathcal{E}(\rho)=(1-p)\rho+p\ket{\psi_{\textrm{ex}}}\bra{\psi_{\textrm{ex}}},$ where $\ket{\psi_{\textrm{ex}}}$ is some external state. For our data the correction obtained by using Eqn.~\ref{eqn:AofE} rather than Eqn.~\ref{eqn:Nielsen formula} is small, reducing the average fidelity from 0.84 to 0.83. 

We obtain a 90\% confidence interval on the value of the average fidelity by a bootstrap analysis similar to the one described in Supplementary Section \ref{errorAnalysis}. We generated 1,000 synthetic data sets, where each data set consists of 20,480 measurement outcomes on each of the four input and output states obtained via tomography. We then run ML on these data sets and compute $F_\mathrm{avg}$. A histogram of the results is shown in Supplementary Fig. \ref{fig:histsF}. Our final result is $F_\mathrm{{avg}}=0.83$ with a 90\% confidence interval of [0.77, 0.86]. 

To estimate the systematic error in our average fidelity calculation which results from truncating the input Hilbert space dimension, we employ the model of the capture process described in Supplementary~Section~\ref{beamsplitterModel}. Our model is a process whose average fidelity can be computed exactly; we find $F_\mathrm{avg}^\mathrm{model}=0.82$. We then simulate the entire experiment, using the model instead of the physical capture process. Specifically, we use the four known input states to generate synthetic data which we perform tomography on. We then send the estimated states through the model process and use the output states to generate synthetic data which we again perform tomography on, before finally calculating $F_\mathrm{avg}$. After 200 repetitions of this procedure we obtain a histogram of average fidelities with mean $\overline{F}_\mathrm{avg}=0.83$. Our estimate of bias is then $\overline{F}_\mathrm{avg}-F_\mathrm{avg}^\mathrm{model}=0.01$, which is small compared to the width of our 90\% confidence interval of [0.77, 0.86]. 

\subsubsection{Classical bound on the average fidelity}

In the main text, it is claimed that the highest possible average fidelity for converting a single qubit state using only classical resources is 2/3. This bound is achieved as follows. Imagine Alice has a qubit prepared in state $\rho_A$, which she wishes to send to Bob through a classical communication channel (e.g., a telephone). If Alice and Bob share a maximally entangled Bell state, then this transmission of quantum information can be achieved via a teleportation protocol\cite{Bennett1993}. However, the use of such an entangled state would constitute a quantum resource and is thus not allowed in this consideration. The best Alice can do then is measure the qubit in some basis, and report the outcome to Bob, who then prepares his qubit in the eigenstate corresponding to Alice's measurement outcome. The quantum operation describing this process is
\begin{equation}
    \mathcal{E}(\rho)=P_0\ket{0}\bra{0}+P_1\ket{1}\bra{1}
\end{equation}
where $P_0 = \bra{0}\rho_A\ket{0}$ is the probability that Alice measures $\ket{0}$, and simlarly for $P_1$. By inserting this process into Eqn.~\ref{eqn:entanglement Fidelity}, and then using Eqn.~\ref{eqn:Nielsen formula}, one obtains an average fidelity of 2/3. Ref.~\cite{Massar1995} considers the more general problem of guessing the state of a qubit given an optimal measurement on an ensemble of $N$ identical copies of that qubit. They find a maximum possible fidelity of $(N+1)/(N+2)$, which equals 2/3 for an ensemble consisting of a single qubit, $N=1$.

\begin{figure}[!ht]
  \centering
    \includegraphics[scale=1]{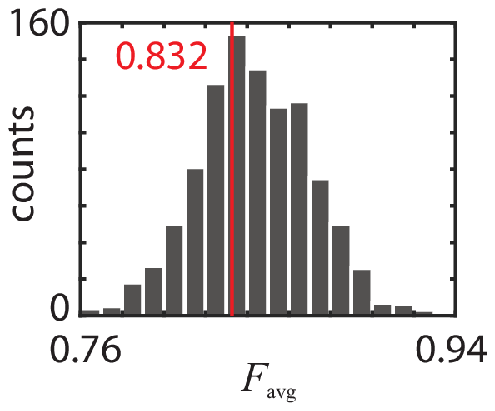}
      \caption{Bootstrapped histograms of the average fidelity. The figure shows a histogram of average fidelities calculated with 1,000 simulated experiments. Each simulated experiment involves generating a synthetic data set of 20,480 Q-function measurements from the known input and output states given in Supplementary Table \ref{tab:densityMatrices}. ML tomography is then performed on each synthetic data set and $F_\mathrm{avg}$ is computed. The red bar indicates the value obtained from the experimental data.}
\label{fig:histsF}
\end{figure}

\begin{table}[!hb]
\caption{Estimated density matrices used to calculate the average fidelity. The states were estimated using 500 iterations of the R$\rho$R algorithm, which yields estimated density matrices of dimension $d=16$. We present the first $3\times3$ elements of these matrices. The input states $\rho_{\textrm{in}}$ were estimated from 20,480 measurements of the input microwave fields after executing the calibration protocol. For the capture protocol, measurements of the mechanical resonator state yield $\rho_{\textrm{m}}$. The states labeled with a `1' (`2') correspond to a vacuum input state (mixed single photon state). Similarly, states labeled with a `3' and `4' correspond to the an input mixed state of a superposition of zero and one photons, with a phase set by $\varphi = 0$ and $\varphi = \pi/2$, respectively.}
\begin{ruledtabular}
\begin{tabular}{c  l  l }
{\bf{Label  }}	&	{\bf{Input state $\rho_{\textrm{in}}$}}	&	{\bf{Mechanical state $\rho_{\textrm{m}}$}} 
\vspace{2mm} \\
% vacuum input
1 &
$\begin{pmatrix}
0.994+0.000i & 0.007 + 0.005i & 0.005 - 0.026i\\
0.007 - 0.005i  & 0.004+0.000i & -0.000 + 0.001i\\
 0.005 + 0.026i & -0.000 - 0.001i & 0.001+0.000i\\
\end{pmatrix}$ &
$\begin{pmatrix}
0.919+0.000i & 0.005 + 0.001i & 0.010 - 0.006i\\
0.005 - 0.001i &  0.0620+0.000i & -0.016 + 0.001i\\
0.010 + 0.006i & -0.016 - 0.001i & 0.017+0.000i\\
\end{pmatrix}$ \vspace{2mm} \\

% photon input
2 &
$\begin{pmatrix}
0.660+0.000i & -0.013 - 0.039i & 0.030 - 0.010i\\
-0.013 + 0.039i & 0.283+0.000i & 0.040 - 0.021i\\
0.030 + 0.010i & 0.040 + 0.021i & 0.042+0.000i\\
\end{pmatrix}$ &
$\begin{pmatrix}
0.636+0.000i & 0.004 - 0.016i & -0.034 - 0.008i\\
0.004 + 0.016i & 0.281+0.000i & 0.021 - 0.012i\\
-0.034 + 0.008i & 0.021 + 0.012i & 0.075+0.000i\\
\end{pmatrix}$ \vspace{2mm}\\

% PHI = 0 deg
3 &
$\begin{pmatrix}
0.826+0.000i & 0.256 - 0.019i &0.020 - 0.003i\\
0.256 + 0.019i & 0.173+0.000i & 0.010 - 0.005i\\
0.020 + 0.003i & 0.010 + 0.005i & 0.001+0.000i\\
\end{pmatrix}$ &
$\begin{pmatrix}
0.763+0.000i & 0.180 - 0.035i & 0.018 - 0.015i\\
0.180 + 0.035i & 0.197+0.000i & 0.024 - 0.011i\\
0.018 + 0.015i & 0.024 + 0.011i & 0.036+0.000i\\
\end{pmatrix}$ \vspace{2mm}\\

% PHI = 90 deg
4 &
$\begin{pmatrix}
0.775+0.000i & 0.0451 + 0.294i & -0.037 - 0.015i\\
0.0451 - 0.294i & 0.217+0.000i & -0.027 + 0.018i\\
-0.037 + 0.015i & -0.027 - 0.018i & 0.006+0.000i\\
\end{pmatrix}$ &
$\begin{pmatrix}
0.759+0.000i & 0.000 + 0.21i & -0.031 - 0.014i\\
0.000 - 0.21i & 0.234+0.000i & 0.019 - 0.017i\\
-0.031 + 0.014i & 0.019 + 0.017i & 0.004+0.000i\\
\end{pmatrix}$ \vspace{2mm}\\

\end{tabular}
\end{ruledtabular}
\label{tab:densityMatrices}
\end{table}

\section{Product disclaimer}
Any mention of commercial products is for information only; it does not imply recommendation or endorsement by NIST.

\end{document}